\documentclass{aa}
\usepackage[varg]{txfonts}
\usepackage{graphicx}
\usepackage{amsmath}
\usepackage{hyperref}
\usepackage{multirow}
\usepackage{lscape}
\usepackage{soul}
\usepackage{longtable}
\usepackage{xcolor}
\usepackage{subcaption}
\usepackage{hyperref}
\newcommand{\kmsmpc}{km~s$^{-1}$Mpc$^{-1}$}
\newcommand\hmpc{$h^{-1}$Mpc}
\newcommand{\kms}{km~s$^{-1}$}

\newcommand{\HI}{\mbox{H\,{\sc i}}}

\newcommand{\h}{$h^{-1}$}
\newcommand{\Msol}{$\textrm{M}_\odot$}
\newcommand{\sbr}[1]{_{\textrm{#1}}} % roman subs
\DeclareUnicodeCharacter{2003}{}
\DeclareUnicodeCharacter{2212}{}
\usepackage[utf8]{inputenc}
\usepackage[T1]{fontenc}
\usepackage{natbib}

\begin{document}

\title{Hidden Vela supercluster revealed by the first hybrid redshift and peculiar velocity reconstruction}

\author{A.M.~Hollinger\inst{1,2}\thanks{amber.hollinger@anu.edu.au}, H.M.~Courtois\inst{1}, R.C.~Kraan-Korteweg\inst{3}, J.~Mould\inst{4,5}
and S.H.A.~Rajohnson\inst{6}}
\institute{
Universit\'e Claude Bernard Lyon 1, IUF, IP2I Lyon, 4 rue Enrico Fermi, 69622 Villeurbanne, France
\and 
 Research School of Astronomy and Astrophysics, Australian National University, Canberra, ACT 2611, Australia
 \and
Department of Astronomy, University of Cape Town, Private Bag X3, Rondebosch 7701, South Africa
\and
Centre for Astrophysics \& Supercomputing, Swinburne University, Hawthorn, VIC 3122, Australia
\and
ARC Centre of Excellence for Dark Matter Particle Physics
\and
INAF Osservatorio Astronomico di Cagliari, Via della Scienza 5, I-09047 Selargius, (CA), Italy
}

\date{Received A\&A March 7th, 2026; Accepted date: June 22nd, 2026}

\abstract {
A large fraction of the extragalactic sky is obscured by foreground dust and stars along the plane of the Milky Way,  leaving a major gap ($\approx 20$\%)  in whole-sky maps of large-scale structures -- an incompleteness that is even more severe for peculiar velocity samples.  This has long limited an unambiguous interpretation of observed cosmic flows and their  connection to the underlying mass-density field. 

We present a new hybrid reconstruction methodology that combines 65,518 galaxy peculiar velocity distances from the CF4++ catalogue with 8283 new galaxy redshifts observed near the southern Galactic plane ($|b| \le 10\degr$) zone of avoidance (ZOA). A major advance is the inclusion of 2176 high-sensitivity, interferometric \HI\ redshifts obtained with the SARAO MeerKAT telescope, which, for the first times provides coverage of the innermost $3\degr$-wide strip of the southern ZOA and to an unprecedented depth.

This hybrid redshift–peculiar velocity approach yields a substantially revised picture of the inferred over-density field in and around the zone of avoidance. In particular, the Vela supercluster emerges as a dominant mass concentration, approaching that of the Shapley concentration and exceeding the mass associated with Laniakea and the Great Attractor region. At a distance of  189 \hmpc , a total mass of $30.6 \times 10^{16}$ M$_\odot$, a characteristic radius of 74 \hmpc, and a double core morphology, Vela dominates the mass budget and gravitational influence of the southern ZOA. These results provide the most complete and dynamically consistent picture to date of the southern ZOA and demonstrate the transformative potential of hybrid reconstruction techniques tailored for the next generation of large-scale surveys.}


\titlerunning{Vela supercluster}
\authorrunning{Hollinger et al.} 
\maketitle
%
%-------------------------------------------------------------------
\section{Introduction}

Roughly a quarter of the extragalactic nearby Universe lies hidden behind the Milky Way, in what is commonly referred to as the zone of avoidance (ZOA). This obscured volume cannot be neglected when searching for potential deviations from the standard cosmological model, $\Lambda$ Cold Dark Matter ($\Lambda$CDM), or from general relativity when studying the local cosmos: ignoring 25\% of the volume risks biasing any inference on large-scale flows or density fields. However, obtaining galaxy distance measurements in this region is notoriously difficult, sometimes impossible, because galaxy images are heavily contaminated by the overwhelming number of foreground stars. This stellar crowding prevents reliable photometric parameter extraction. Moreover, corrections for Galactic foreground extinction add further uncertainties. Even in the $K$-band, extinction is non-negligible close to the dust equator. An error in extinction of $A_K \sim 0.1$ mag translates into a 200 \kms\ uncertainty in peculiar velocities, which precludes its use as a standard distance indicator.

Despite these obstacles, galaxies in the ZOA can still be detected and positioned in redshift space with radio telescopes through the 21\,cm line emission of neutral hydrogen gas \cite[e.g.,][]{HIZOA-2016}. With the advent of the Square Kilometre Array (SKA) and its precursor facilities, including the South African MeerKAT radio telescope and the Australian Square Kilometre Array Pathfinder (ASKAP), systematic interferometric \HI\ surveys are being pursued, such as: WALLABY \citep{Koribalski-2020}, DINGO \citep{Meyer-2009}, MIGHTEE-HI \citep{Maddox-2021}, FORNAX \citep{Serra-2016,Serra-2023}, and LADUMA \citep{Blyth-2016}, that allow significantly increased sensitivity, instantaneous redshift coverage, as well as positional accuracy compared to the earlier single-dish \HI\ surveys. They anchor these galaxies within the Hubble flow. After many years of data collection in South Africa, this work presents the first reconstruction of the obscured region using several thousand galaxies, and it marks the first time that a peculiar-velocity-based reconstruction \citep{Courtois2012, Courtois2023, Courtois2025} has been completed using a dataset that includes galaxies for which only redshifts, not distances, are available.

Developing and testing such a hybrid redshift and peculiar-velocity methodology is an essential preparation for the next generation of large surveys, which will simultaneously deliver both types of data over vast cosmic volumes. Projects such as the Dark Energy Spectroscopic Instrument (DESI)\footnote {\url{www.legacysurvey.org}}, the 4MOST Hemisphere Survey of the Nearby Universe (4HS) \citep{Taylor_4HS}, one of the flagship extragalactic surveys of the 4-metre Multi-Object Spectroscopic Telescope (4MOST), and WALLABY are already beginning to assemble matched samples of redshift and velocity information, dramatically expanding the parameter space of future reconstructions. Establishing robust techniques that can integrate galaxies with full distance information together with those for which only redshifts are available is therefore a crucial step. It ensures that forthcoming datasets can be exploited to their full potential, leading to more accurate maps of the local large-scale structure and more stringent tests of the $\Lambda$CDM model. Within the $\Lambda$CDM paradigm, large-scale structures emerge from the gravitational growth of initial density fluctuations dominated by CDM, while cosmic acceleration is attributed to a cosmological constant $\Lambda$ \citep{planck_2020}.

This paper is organized as follows. In Section \ref{sec:datasets} we summarize the datasets and methodology used to integrate the ZOA galaxies into the CF4++ catalogues within the Local Universe, adopted in this paper as $z < 0.1$ (300 \hmpc). We
present and discuss our results in Section \ref{sec:results}, followed by our conclusions in Section \ref{sec:conclusion}.

\section {Integration of ZOA velocity datasets to CF4++ database}\label{sec:datasets}

In this section, we outline the different datasets compiled for this study. Our investigation concentrates on the southern ZOA, which hosts some of the most prominent over- and under-densities in the nearby Universe, including the Local Void \citep[LV: ][]{Tully-2008,Kraan-2008}, the Ophiuchus Supercluster \citep[Oph:][]{Hasegawa-2000,Wakamatsu-2005}, the Great Attractor \citep[GA:][]{Dressler-1987,Kraan-1996,Woudt2004}, and the Vela Supercluster \citep[VSCL: ][]{Kraan-2017}. To better characterize these structures and the flow fields they induce, numerous extensive, long-term galaxy surveys have been conducted. However, the paucity of ZOA data, particularly at higher redshifts where major contributions to the residual bulk flow are thought to arise, has remained a key concern. 
For instance, systematic \HI\ ZOA-surveys were limited to $z = 0.04$ with little sensitivity above $z=0.025$ (see {\sl Dataset C1}), while optical spectroscopic coverage is very patchy, and the only systematic redshift survey 2MRS (the 2MASS Redshift Surveys  (\citep{Huchra2012, Macri2019}) has little sensitivity beyond $z=0.05$ and does not probe the inner ZOA ($|b| < 5 \degr$) at all.
Recently, dedicated MeerKAT \HI\ surveys have delivered a substantial breakthrough by probing the innermost, fully opaque ZOA to unprecedented depths in sensitivity, angular resolution, and redshift range (z < 0.085), in addition to a number of spectroscopical surveys at intermediate ZOA latitudes. 

\begin{figure}[h!]  
\centering
\includegraphics[width=\linewidth,angle=-0]{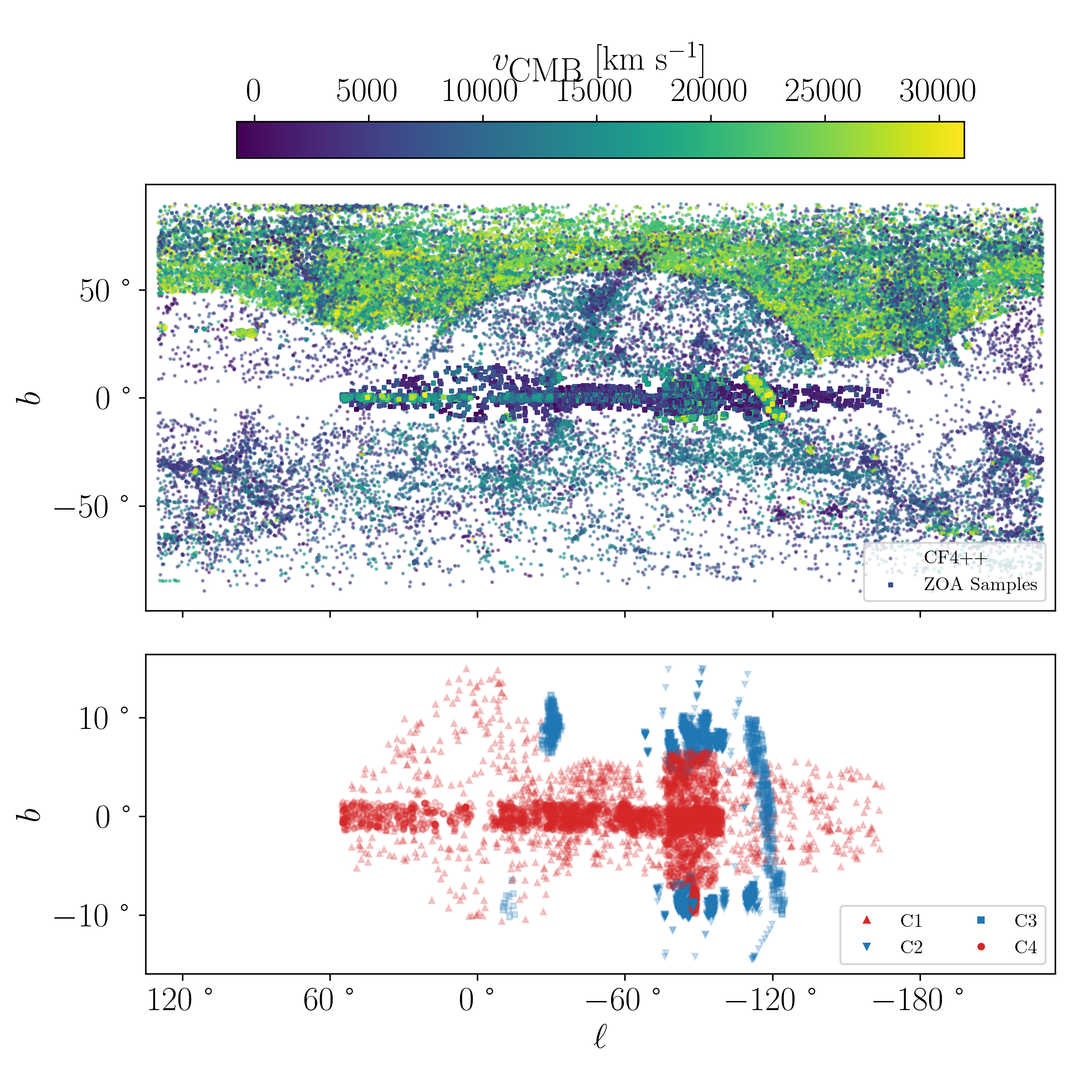}
\caption{Galactic $\ell,b$ distribution of: (top) the CF4++ZOA samples along with their measured redshifts; and (bottom) the 8283 ZOA C1-C4 samples introduced in this work (red for \HI\ and blue for optical redshifts). Note the unprecedented dense coverage provided by the seven datasets obtained with SARAO MeerKAT telescope (red circles) along the innermost, nearly fully opaque, ZOA region.}
\label{fig:ZOA_vs_CF4}
\end{figure}

The top panel of Fig.~\ref{fig:ZOA_vs_CF4} displays the distribution in Galactic coordinates on the sky of the CF4++ together with the ZOA redshift measurements presented below, while the bottom panel provides a closer view of recent ZOA redshifts. The ZOA datasets are presented in four categories, a summary of which can be found in Table \ref{tab:catalog_summary}. The different symbols point to their respective sources and observations. The red symbols denote 21cm redshifts with the triangles originating from the Parkes \HI\ ZOA surveys (see {\sl Dataset C1}), and the circles the recent MeerKAT \HI\ detections that reach to redshift $z=0.08$ described in {\sl Dataset C4}). The blue symbols represent to optical redshifts. These result from dedicated redshift campaigns in the Vela region using a variety of telescopes and instruments (triangles; see {\sl Dataset C2}, and other southern ZOA regions (squares; {\sl Dataset C3}). The optical redshifts will be published at a later stage (Kraan-Korteweg\ \& Cluver, in prep.).

\begin{table*}
\centering
\caption{Summary of the catalogues contributing to the hybrid ZOA compilation.}
\label{tab:catalog_summary}
\begin{tabular}{p{1.2cm} p{1.6cm} p{3.2cm} p{.5cm} p{3.2cm} p{4.6cm}}
\hline
Catalogue & Detection Method & Region & $N$ & Survey / Instrument & References \\
\hline

C1 & \HI\  & Southern ZOA, Northern and Galactic Bulge extensions & 1096 & Parkes Multibeam HIZOA survey & \citet{HIZOA-2016}, \citet{HIZOANE_2005}, \citet{Kraan-2008} \\

C2 & Optical spectroscopy & Vela Supercluster region & 4507 & SpUpNIC, MOS, 2dF+AAOmega, 6dF-MOS, Optopus & \citet{Kraan-2015}, \citet{Kraan-2017}, \citet{Hatamkhani-2024}\\ 

C3 & Optical spectroscopy & Puppis and Scorpius regions & 669 & 6dF &  \citet{Jarrett2MASX-2000} \\

C4 & \HI\  & Southern Galactic Plane and Vela extensions & 2825 & MeerKAT SMGPS, Vela-HI, and pilot surveys & \citet{steyn2023}, \citet{Goedhart-2024}, \citet{Kurapati-2024}, \citet{Rajohnson-2024a}, \citet{Rajohnson-2024b}, \citet{Steyn2024GA} \\

\hline
\end{tabular}
\end{table*}

\paragraph{Dataset C1 (N=1096)}:\\
C1 contains \HI\ detected galaxies from the Parkes Multibeam ZOA survey (r.m.s. sensitivity of 6~mJy~beam$^{-1}$), which was the first systematic \HI\ survey of the southern ZOA specifically designed to trace obscured large-scale structures. HIZOA is composed of three components: the primary HIZOA catalogue \citep[HIZOA-S][]{HIZOA-2016}, a northern extension \citep[HIZOA-NE][]{HIZOANE_2005} covering $52\degr > \ell > 196\degr$ for $|b| < 5\degr$, and a third extension targeting the Galactic Bulge (HIZOA-GB) (see \cite{Kraan-2008}, Kraan-Korteweg et al. in prep.), where stellar crowding and extinction extend to higher latitudes, covering latitudes $b = \pm10\degr$ for $332\degr \le \ell \le 36\degr$, and $b=+15\degr$ for $348\degr \le \ell \le 20\degr$ (as indicated by red triangles around Galactic Bulge in bottom panel of  Fig.~\ref{fig:ZOA_vs_CF4}). All three subsets were obtained with an identical instrumental configuration. The combined HIZOA survey sample consists of  1096 galaxies detected out to c$z$ $\sim 12{,}700$\kms. Note, however, that the number of detections declines markedly beyond the distance range of the GA. Owing to the relatively large positional uncertainty (4 arc min), cross-identification of sources--if visible in the ZOA--is particularly challenging.

\paragraph{Dataset C2 (N=4507)}:\\
Optical spectroscopy in the Vela Supercluster region ($\ell = 272.5\degr \pm 20\degr, b = 0\degr \le \pm 10\degr$). The extended Vela ZOA region has been a focus of interest for some time, as prominent, distant large-scale walls appear to vanish behind the ZOA, and early optical ZOA galaxy surveys had already revealed marked over-densities at similar distances \citep[e.g.,][]{Kosma-2012}. This motivated an extensive spectroscopic program across the ZOA in 2012, targeting galaxies in areas where extinction and stellar confusion are still moderate. Redshift measurements were acquired using SpUpNIC on the 1.9m telescope (SAAO), the multi-object spectrograph (MOS) on the Southern African Large Telescope \citep[SALT;][]{Kraan-2015} for candidate rich clusters, and -- representing a substantial step forward -- 25 fields observed with the 2dF+AAOmega spectrograph (392 fibres) on the 3.9m telescope of the Australian Astronomical Observatory (AAO). The latter yielded 4100 new, unique redshifts. When these are combined with earlier (unpublished) Vela redshift data obtained with the 6dF-MOS on the 1.2m Schmidt telescope at the AAO and with Optopus on the 3.6m telescope at ESO, the sample reaches 4432 redshifts in total \citep[see][]{Kraan-2017}. The recently reported 75 additional SALT redshifts in a rich VSCL cluster \citep{Hatamkhani-2024} contribute an additional 75 galaxies.

\paragraph{Dataset C3 (N=669)}:\\
These measurements correspond to redshifts obtained via optical spectroscopy using the 6dF instrument on the 1.2m UK Schmidt Telescope at the AAO, collected during the early testing phase of the 6dF project. The 6dF pointings were situated within the ZOA ($|b| < 10\degr$) and thus lay outside the main 6dF survey \citep{Jones6dF-2009}, from which the ZOA was excluded. Two supplementary programs were carried out: one targeted fields in the Puppis constellation -- specifically, a strip spanning the full ZOA for $|b| < 10\degr$ and centred at $\ell \sim 245\degr$ -- yielding 388 redshifts for 2MASX galaxies \citep{Jarrett2MASX-2000}, four of these are listed in HIZOA, hence 384 new redshifts. The other focused on the Scorpius region near $\ell \sim 340\degr$, providing an additional 285 redshifts (from optically identified and 2MASX galaxies), the majority of which lie at comparatively higher ZOA latitudes.

\paragraph{Dataset C4 (N=2825)}:\\
This dataset summarizes the systematic \HI\ surveys that were undertaken in the southern ZOA. The observations were performed with the MeerKAT L-band receiver (856$-$1712MHz), and \HI\ data were extracted for the range 1308$-$1430MHz ($z < 0.085 $ or $cz < 25\,000$\,\kms), the volume within which RFI is minimal, while encompassing the full distance range relevant to (residual) bulk flows.

A first MeerKAT16 Science Verification experiment using the 16-dish sub-array setup and the 32k ROACH correlator was carried out in May 2018 as a pilot study. It targeted six slightly overlapping pointings situated in a region of high galaxy density in Vela (around $(\ell,b) \sim (282\degr, -8.3\degr$). The 3-hour observations achieved an r.m.s. noise level of 1.2~mJy~beam$^{-1}$ and an excellent velocity resolution of 5.6 \kms\ at $z = 0$. A thorough examination of the resulting data products \citep{steyn2023} yielded 119 robust detections (excluding 36 tentative galaxy candidates), 91 of which had no previously known redshifts in dataset C2. The pilot survey thereby demonstrated that typical spiral galaxies can be readily detected out to the highest redshifts probed.

The success resulted in a proposal (led by RCKK) to exploit the then still ongoing SARAO MeerKAT Galactic Plane Survey \citep[SMGPS][]{Goedhart-2024}, a narrow survey of $|b| = 3\degr$  along the Galactic Plane ($\sim~900$ interleaving pointings of about 1~hr  each aimed at continuum studies of the Galaxy) for a systematic \HI\ survey. The resulting \HI\ data cubes led to an r.m.s. sensitivity of $0.3-0.6$~mJy~beam$^{-1}$, albeit with low velocity resolution (44~\kms\ at  $z = 0$). The overall SMGPS \HI\ survey covered the longitude range from $260\degr <  \ell < 55\degr$ for $z$ < 0.085 (c$z$ < 25,000\kms). All data were reduced in a similar manner, \citep[see e.g.,][for details on the data reduction and source extraction process]{Rajohnson-2024a}. The analysis was divided into several sub-projects, each concentrating on specific--known or suspected--large-scale structures. 

\cite{Kurapati-2024} investigated the extended Local Void (LV), covering Galactic longitudes from $55\degr > \ell > 332\degr$ and constrained to c$z$ < 7500\kms\ to encompass the full LV volume. They detected 291 galaxies in this region. These new detections substantially improved the mapping of the LV boundaries at these Galactic latitudes. A subsequent study analysed the higher redshift range to uncover potential connections to  the Ophiuchus supercluster, unveiling an additional 431 galaxies (for early results, see \cite{LouwIAU-2026,LouwMSc-2026}, increasing the total to 722 \HI\ detections (30 in common with HIZOA). 
The adjoining SMGPS longitude range, $332 \degr > \ell > 302\degr$, was surveyed to evaluate the prominence of the GA Wall as it cuts across the ZOA. Of the 477 galaxies detected there (42 already catalogued in HIZOA), nearly half trace an exceptionally pronounced wall at the distance of the GA \citep{Steyn2024GA}. The subsequent Crux SMGPS region was examined later to bridge the gap between GA and Vela, $302 \degr > \ell > 290\degr$; the final assessment is still underway \citep[see][for preliminary results]{SteynIAU-2026}. In this sector, 289 galaxies were detected (12 previously in HIZOA), revealing a so far unknown over-density that may well be associated with the overall GA structure.
In the SMGPS-VSCL longitude range, $290\degr > \ell > 260\degr$, the highest number of \HI\ detections was obtained (843 galaxies; 39 listed in HIZOA), tracing multiple large-scale structures at different distances, and a prominent peak at the distance of the Vela SCL \citep{Rajohnson-2024a}. 

The combined SMGPS \HI\ detections--eliminating duplicates in overlapping regions--yields a total of 2288 \HI\ detections (112 of which are in HIZOA) hence 2176 are new redshift measurements. The total might still change slightly once the full SMGPS analysis has been completed.

A further \HI\ survey has been conducted with MeerKAT \citep[Vela-HI; ][]{Rajohnson-2024b} to extend the SMGPS to higher latitudes (about $-6.7\degr$ to $+6.2\degr$) over the longitude range $263\degr  \le  \ell \le 284\degr$ to trace the suspected connection of the Vela Walls seen in the optical spectroscopic surveys (see dataset C1), with those identified in the SMGPS \citep{Rajohnson-2024a}. The setup was similar, although the r.m.s. sensitivity was slightly lower, while the beam was slightly larger \citep[see][for details]{Rajohnson-2024b}. A total of 719 galaxies were identified, of these 16 had an overlap with the aforementioned Vela-SMGPS, 79 of the remainder had HIZOA counterparts (C1), 66 had optical velocities (given in C2), leaving a further 558 new ZOA redshifts after removing all overlaps. 

Taking the four datasets described above of `southern' ZOA redshifts, from a total of 9097 unique redshifts, 8283 (using our cut-off at  30000\kms) were added to the CF4++ catalogue of 65,518 galaxy distances. The coverage is displayed in Galactic coordinates in Fig. \ref{fig:ZOA_vs_CF4}, where the top panel presents the merged CF4++ZOA survey. The sample now comprises 73,801
individual galaxies. The bottom panel shows a close-up of southern ZOA data (red for \HI\ and blue for optical redshifts). The power of the MeerKAT telescope in penetrating the innermost ZOA (C4: red circles), where little was known beforehand, is transformative.

%fig1

\begin{figure}[h!]  
\centering
\includegraphics[width=\linewidth,angle=-0]{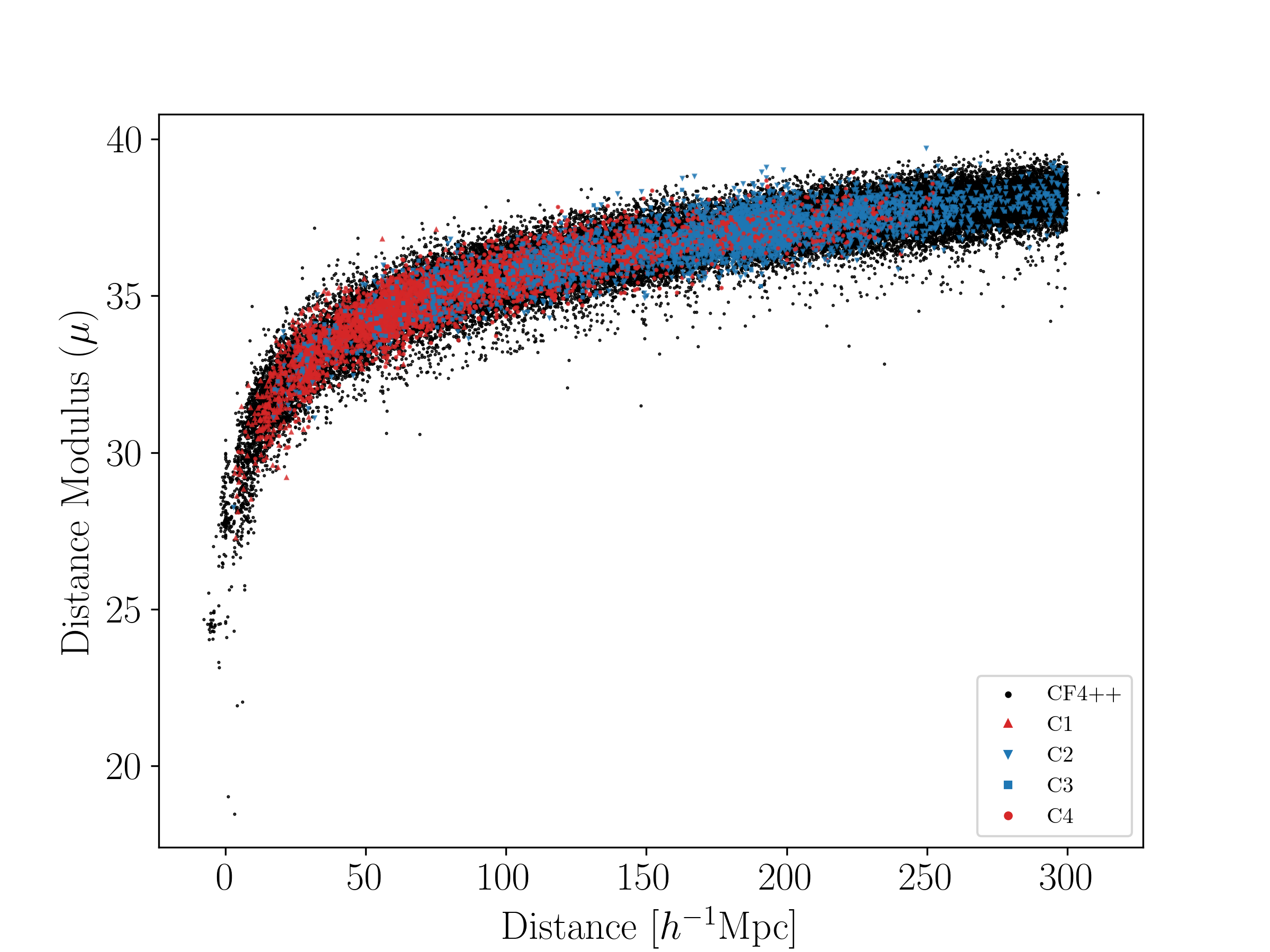}
\caption{Comparison of the CF4++ (black) distance moduli as a function of $V_{\rm CMB}$/100 compared to the pseudo-distance moduli adopted for the 8283 galaxies in the ZOA datasets (red for \HI\ and blue for optical redshifts).}
\label{fig:ZOA_pos}
\end{figure}

To use this ensemble of ZOA redshifts in the CMB rest frame, we adopt the value of $H_0=74.6$ \kmsmpc, consistent with CF4++, and derive estimated luminosity distances and distance moduli. Since this relation is intrinsically scatter-free, we introduce realistic uncertainties by perturbing the distance moduli with Gaussian noise. For each radial bin, we calculate the standard deviation of the corresponding CF4++ distance moduli and use it as the width of the Gaussian perturbation applied to the ZOA sample. The coloured points in Fig.~\ref{fig:ZOA_pos} show the resulting pseudo-distance moduli for our ZOA sample, compared to those of CF4++ in black. The distance modulus uncertainties are assigned in a similar manner, with hard limits imposed to ensure consistency with CF4++. In particular, these limits ensure that an object always satisfies the minimum uncertainties observed at different redshifts within the catalogue. That is to say, for our ZOA sample we neglect peculiar velocities themselves, but not the associated measurement uncertainties. The perturbed distance moduli and their corresponding uncertainties are then incorporated into the reconstruction.

%fig2

\subsection{Test datasets}\label{sec:test-datasets}
To verify that the hybrid reconstruction scheme does not produce artificial large-scale structures, we performed a series of tests grouped into three categories: 
\begin{enumerate}
    \item {\it ZOA}: distance moduli and peculiar velocities for the ZOA galaxies are derived from the CF4++ data, to confirm that our new reconstruction introduces only minimal changes compared to our previous reconstruction.
    \begin{enumerate}
        \item {\it -- grid}: the peculiar velocities and their uncertainties are  assigned based on the CF4++ reconstruction grid. This results in distance moduli and uncertainties that exhibit significantly less scatter than the final ZOA C1--C4 galaxy samples.
        \item  {\it -- nearest}: these quantities are assigned using the values of the nearest galaxy in the CF4++ catalogue directly.  
    \end{enumerate}
    \item {\it MDPL2}: galaxy positions and peculiar velocities are taken directly from a mock catalogue drawn from the MultiDark Planck 2 (MDPL2) simulation (see Appendix \ref{app:tests}  for more details). These test cases were performed to demonstrate that the reconstruction will not find non-existent structure and that we can test the robustness of the final results.
    \begin{enumerate}
        \item {\it -- area}: reproduces the radial selection function of the C1--C4 catalogues, while applying approximate Galactic latitude limits of $7.5\degr$ and $12\degr$ depending on Galactic longitude, see Fig. \ref{fig:ZOA_vs_CF4}. In this case, Galactic longitudes are assigned randomly.
        \item {\it --  matching}: in contrast,  reproduces the radial selection function and matches the distributions of Galactic coordinates and object counts.
    \end{enumerate}
    \item {\it Random}: this variant uses a catalogue of randomly generated positions and velocities. Similar to the MDPL2 cases, this ensures that assigning non-existent galaxies will not artificially introduce structure.
    \begin{enumerate}
        \item {\it -- area}: same as before. 
        \item {\it -- matching}: same as before.
        \item {\it -- fully-random}: galaxy positions and distances are drawn entirely at random.

    \end{enumerate}
       
\end{enumerate}

 \begin{figure}[h!]
 \centering
\includegraphics[width=\linewidth,angle=-0]{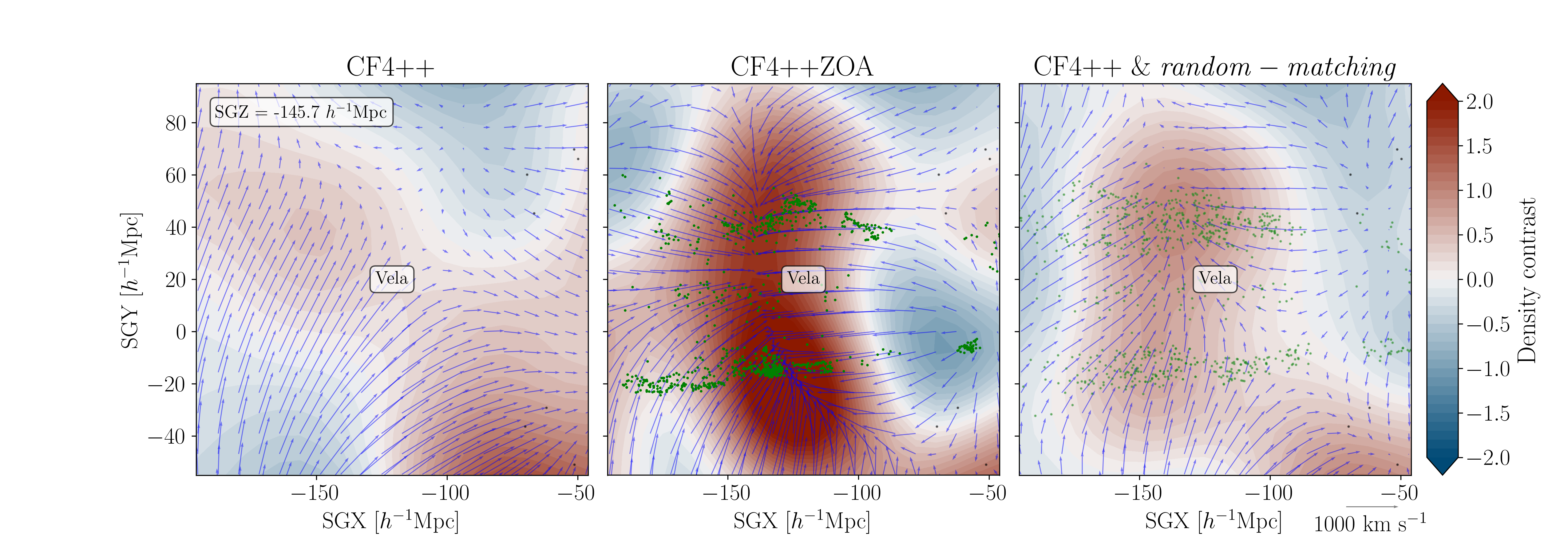} \\
  \caption{Differences in the Vela cosmography for the CF4++ (left), CF4++ZOA (middle), and  CF4++ \& \textit{random-matching} (right) density reconstructions.  The green points shown in the middle and right panels indicate the positions of the artificial galaxies added in each respective case.    }
\label{fig:random}
\end{figure}

Several tests were performed to check the robustness of the reconstructions. For example, Fig. \ref{fig:random} shows the Vela supercluster region for CF4++, CF4++ZOA, and CF4++ \& \textit{random--matching},
the latter demonstrates the impact of adding an artificial dataset created with the same number of galaxies, distributed over similar Galactic coordinates and depths as the ZOA observed galaxies, but assigned random peculiar velocities. Compared to the CF4++ density field both the ZOA and \textit{random--matching} cases introduce a level of change to the local cosmography. However, the random data does not introduce significantly larger density peaks than the original CF4++, nor significant changes to the overall velocity field when compared to the CF4++ZOA case. This clearly demonstrates that the reconstruction algorithm interprets this as white noise and that there is no significant correlation signal in this artificial dataset. Similarly, all seven tests confirm that the hybrid reconstruction does not introduce spurious artifacts in either the density or velocity fields, nor generate artificial signals to the bulk flow (see Appendix \ref{app:tests} for further details).

In addition to these tests, an independent study using both the Tully-Fisher distances and the redshifts for a subset of the ZOA galaxies concludes that the recovered velocity field in the Great Attractor region of the ZOA, with fractional peculiar velocities of $\Delta v_{pec}/cz \leq 0.1$, allows for reconstruction of the region without recourse to redshift-independent distances \cite{MouldZOA2026}.

\section{Results}\label{sec:results}

In this section, we present an updated reconstruction of the Local Universe's velocity and total matter (comprising both dark and luminous components) over-density fields, based on the hybrid model incorporating the ZOA redshift data, and compare it with the previous CF4++ reconstruction \citep{Courtois2025}. The newly presented velocity fields are derived using a self-consistent approach, whereby the inferred field at any given point depends non-locally on the entire dataset. The methodology employed for these calculations aligns with the procedure detailed in Section 4 of \cite{Courtois2023}; for a more comprehensive explanation, we refer the reader to that paper.

\subsection{Notable changes in reconstructed density and velocity fields}

\begin{figure*}[h!]
 \centering
 \includegraphics[width=\linewidth,angle=-0]{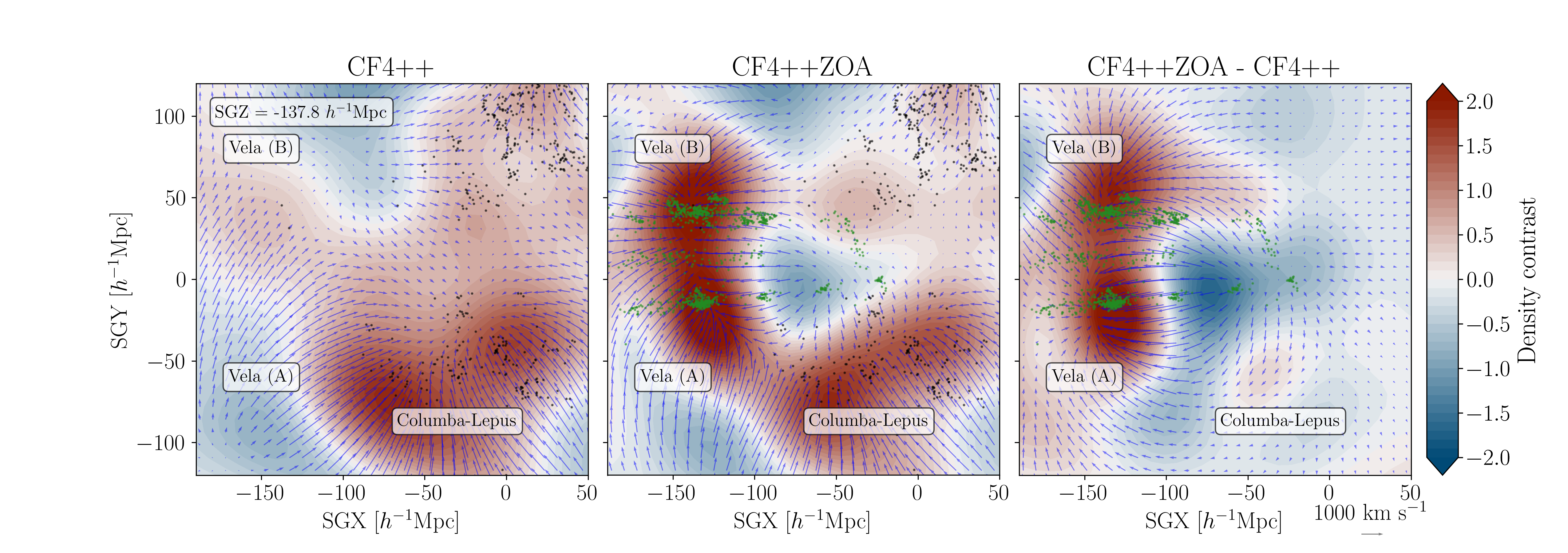} \\
  \caption{Reconstructed matter density field, averaged over a total width of 23.4 \h Mpc in the SGZ plane of the superclusters' regions that demonstrate the biggest changes with the addition of the ZOA datasets: Columba-Lepus and Vela. The left panel displays results derived from the CF4++ dataset, while the middle panel uses the CF4++ZOA dataset, and the right panel shows the difference between the two reconstructions. Overlaid is the mean velocity field. Black dots represent CF4++ galaxies, and the green dots correspond to newly added ZOA data points located within the averaged region.}
 \label{fig:Velachanges}
\end{figure*}

Substantial changes in the reconstructions based on the hybrid dataset compared to the original CF4++ reconstruction were revealed for the Lepus-Columba and the Vela supercluster, while most other previously identified over-densities in CF4++ (see Appendix \ref{app:C}) show only minor variations (at levels below $|\delta|<0.2$). 
Interestingly, the latter include the Local Void, the Great Attractor, and the Ophiuchus supercluster, a significant fraction of which is located within the ZOA
(see Fig.~\ref{fig:nochange}).

Here, we focus on the description of the over-densities whose reconstructed full matter density and velocity fields are most significantly affected by the inclusion of the new data. Figure~\ref{fig:Velachanges} shows an SGZ slice averaged over a width of 23.4~\hmpc\ through the Columba-Lepus and Vela regions, showing, from left to right, the original CF4++ dataset, the CF4++ZOA dataset, and the difference between them.
The CF4++ galaxies are marked by black dots, while the newly incorporated ZOA data points are shown as green dots. The changes in the shape, density, and extent of these structures due to the addition of the ZOA datasets are significant and dramatically alters the local cosmography.  

Columba-Lepus is not located in the ZOA itself, but lies below the Galactic Plane at a lower distance compared to Vela. While previously recognized as a conglomeration of clusters (see the two density concentrations in the left panel), and proposed as a possible supercluster \citep[see Fig.4 in][]{Pomarede2020}, it now emerges as a strong oblong, elongated structure, more like a ridge, that forms part of a boundary of a previously unknown cosmic void extending visibly into the ZOA at SGX,SGY=(75; 0) \hmpc. The highest density core is Columba rather than Lepus. 

Vela is significantly more prominent than in the previous reconstruction using the CF4++ dataset. It reveals two distinct, prominent cores embedded in a much larger structure. While this could be influenced by the distribution in the ZOA redshift dataset, a number of galaxies are also observed between these two density peaks (see middle panel of Fig.~\ref{fig:Velachanges}). Moreover, such an effect has not been observed in other ZOA structures, such as the Local Void, Great Attractor, and Ophiuchus supercluster, where the ample ZOA data resulted in marginal changes to the density field only; see Fig.~\ref{fig:nochange}. 

Near Vela, the velocity field is overall drastically modified and exhibits clear infall and back-fall patterns around the extended Vela supercluster. The two Vela supercluster cores are located at supergalactic coordinates (SGX,SGY,SGZ) =  $(-121.1,-19.5,-144.5)$ \hmpc\ for Core {\it a} and $(-136.7,50.8,-121.1)$ \hmpc\ for Core {\it b}. The reconstructed velocity field suggests that Core {\it a} is the dominant component.

The two cores are separated by 75 \hmpc. Core {\it a} is slightly more massive with $14.4 (3.7)\times10^{16}$  M$_\odot$, while Core {\it b} has $13.7 (3.3)\times10^{16}$  M$_\odot$. Just as in Table~\ref{tab:volume-mass}, the mass in brackets is the central mass enclosed within the iso-contour $\delta = 1.5$. The gravitational escape velocity from the cores at mid-separation distance is about 5700  \kms\. This is much larger than the 2760 \kms\ Hubble expansion velocity at mid-separation distance between the cores. This implies that the internal structure of Vela is still collapsing in co-moving coordinates.

\begin{figure}[h!]
\includegraphics[width=\linewidth]{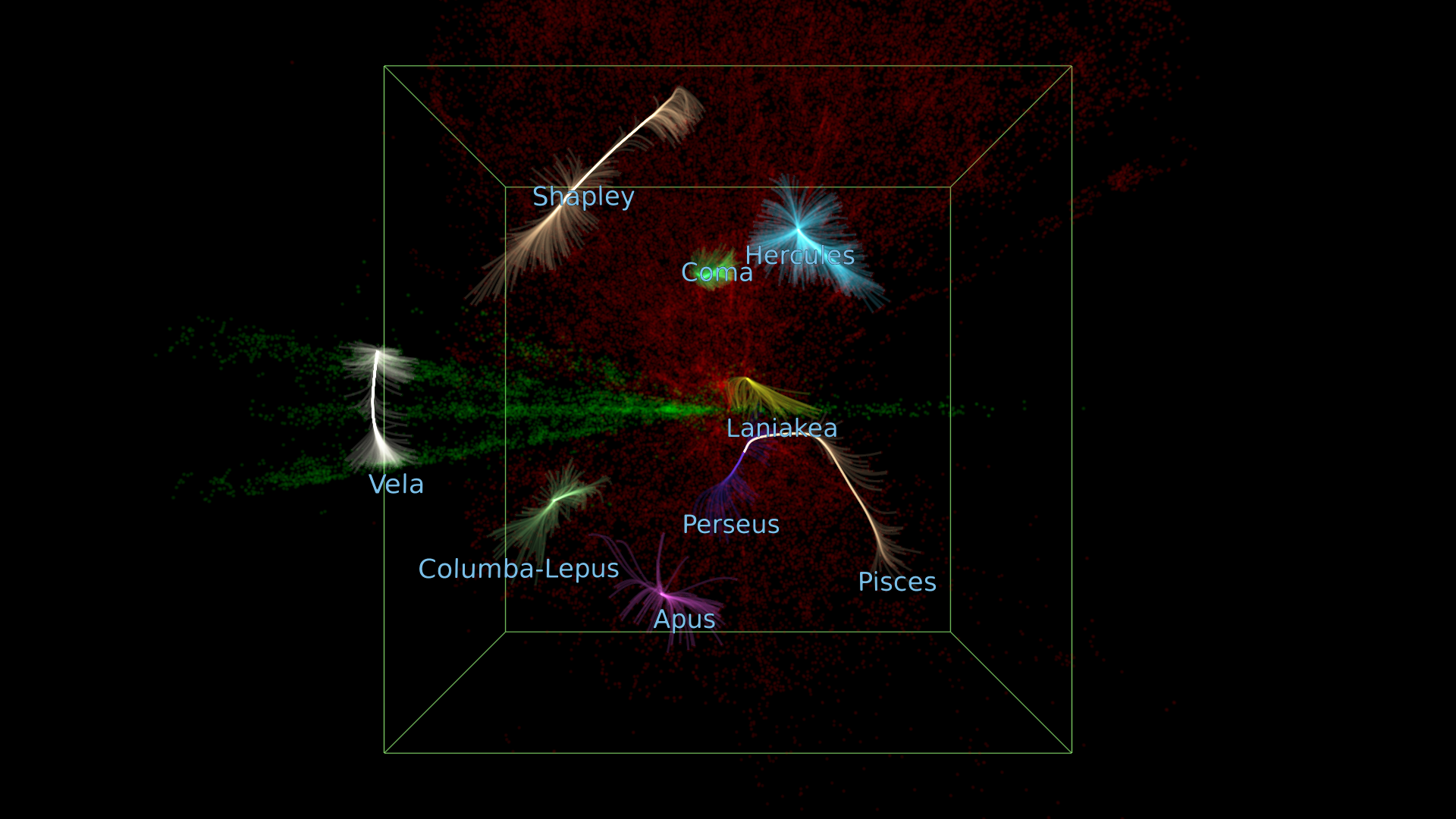}
    \caption{Local superclusters reconstructed using the CF4++ dataset (red dots) expanded by the ZOA dataset (green dots). The green cube provides the 300 \hmpc\ scale length and delineates the orientation primarily showing the SGX-SGY plane with +SGZ pointing forward. Streamlines of the velocity field are coloured according to their gravitational watershed membership. The Vela supercluster emerges as a large single entity gravitational watershed on the left of the image, crossing the ZOA. An associated movie version of this figure is available online.}
    \label{fig:superclusters}
\end{figure}

\begin{figure}[h!]
\includegraphics[width=\linewidth]{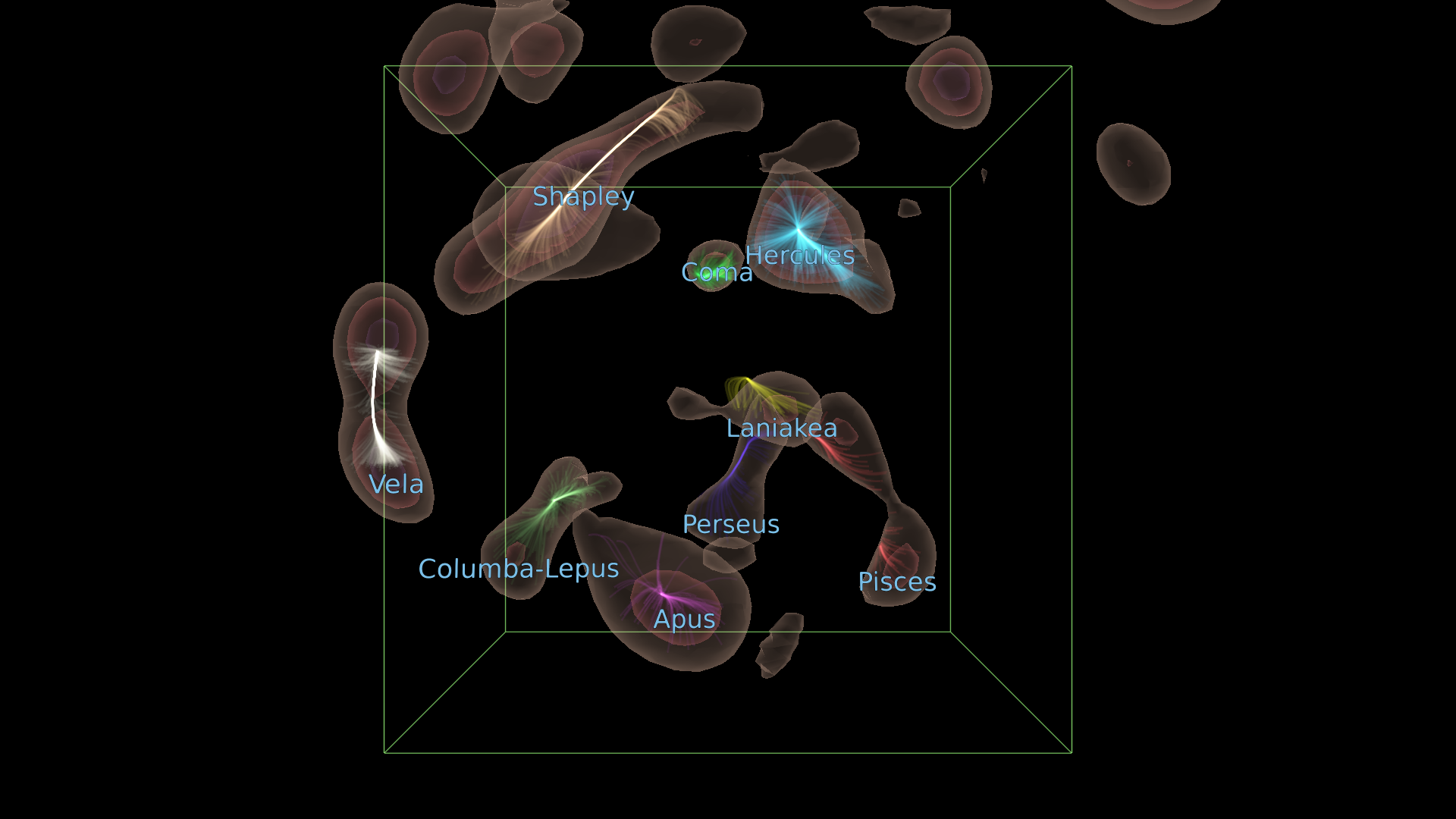}
    \caption{Same streamlines and layout as in Fig. \ref{fig:superclusters}, but with the addition of the  iso-surfaces of the reconstructed density field in the Local Universe. The threshold delta field values for each iso-contour is 1.5, 2 and 2.3. The Vela supercluster extends well beyond the Galactic plane with its internal structure showing a double core. An associated movie version of this figure is available online.}
    \label{fig:superclusters_iso}
\end{figure}

In Fig.~\ref{fig:superclusters}, the reconstructed Vela supercluster is displayed together with other major superclusters of the Local Universe recovered from the CF4++ peculiar velocities reconstructions (Shapley, Hercules, Columba-Lepus, Apus, Pisces, Perseus, Laniakea, Coma). A streamline of the velocity field represents the spatial integral of the velocity field at a given time, tracing the instantaneous direction of motion through space rather than the time evolution of individual particles. Watersheds are then defined as the regions that collect all streamlines converging toward the same attractor. When the flow crosses a cosmic void or becomes divergent, streamlines separate, and these regions of divergence define the ridge lines that form the boundaries between neighbouring watersheds. Figures~\ref{fig:superclusters} and ~\ref{fig:superclusters_iso} show that Vela indeed is a single gravitational watershed entity, with a double core that crosses and extends beyond the Galactic plane. Both Figures are accompanied by online movies.

%fig5

Iso-contours of the density field provide a powerful tool to identify both the shape and the volume that define the boundaries of a supercluster. By selecting fixed thresholds of the density contrast, $\delta$, one can delineate regions of coherent over-density and distinguish them from the surrounding large-scale structure. In our analysis (see Fig.~\ref{fig:superclusters_iso}) we show the iso-contours corresponding to  $\delta$ = 1.5, 2 and 2.3,  where $\delta = 0$ corresponds to the cosmic mean density. Positive values of $\delta$ trace over-dense structures, while negative values would identify cosmic voids. The iso-contour at $\delta$ = 1.5 encloses a large, contiguous volume associated with the Vela supercluster, while the iso-contours at  $\delta$ = 2 and $\delta$ = 2.3 reveal the double core structure of Vela. Moreover, Fig.~\ref{fig:superclusters_iso} delineates its spatial extent and total enclosed volume geometrically, providing further evidence that the Vela structure extends well beyond the Galactic plane.

\subsection{Dynamical masses of eight nearby superclusters}

\begin{figure}[h!] 
\centering
\includegraphics[width=\linewidth,angle=-0]{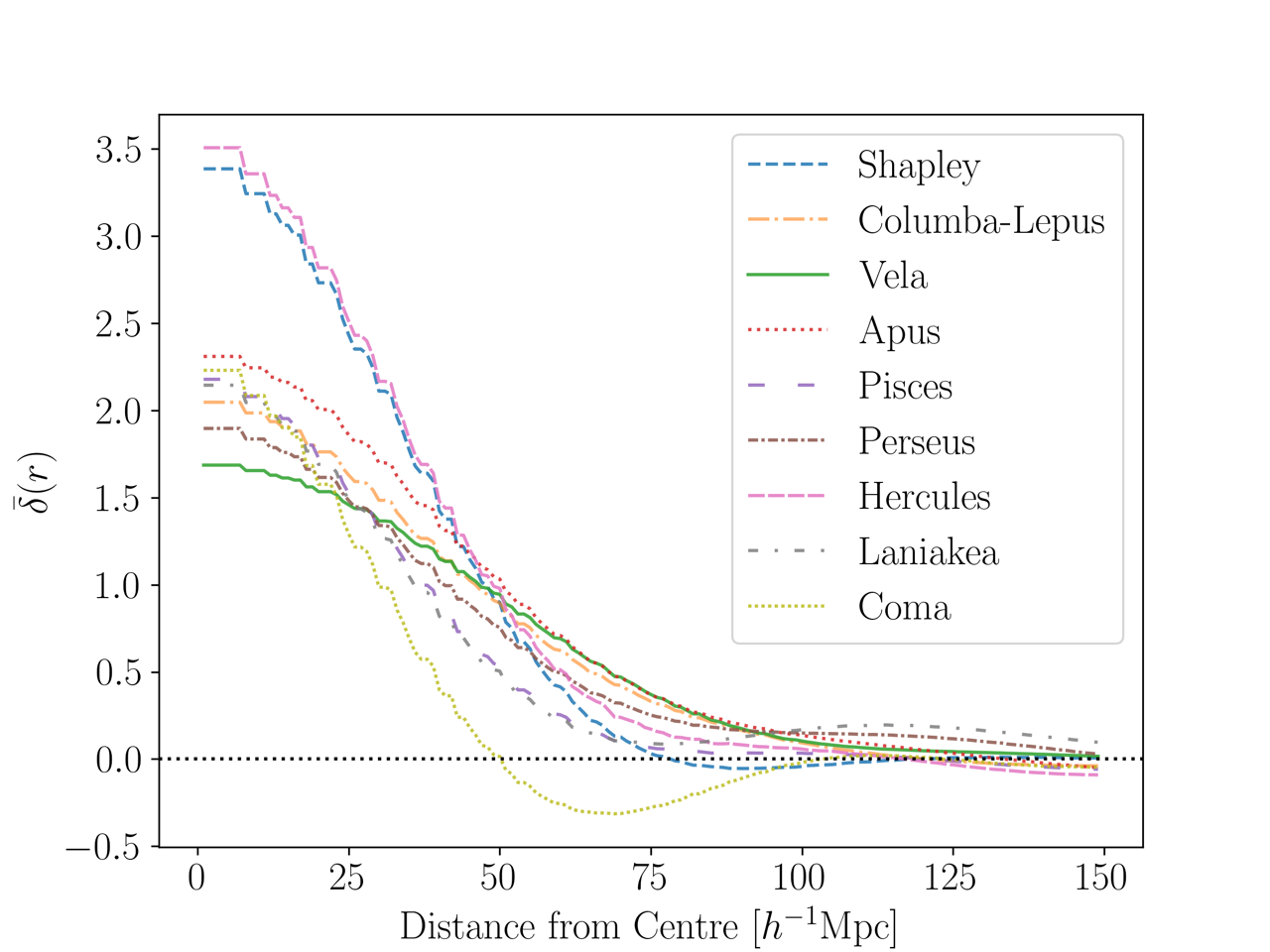}
\caption{Mean density profile  of the local supercluster watersheds as a function of distance from their centre. The majority of these superclusters attain the average cosmic density at distances of $\sim$70 \h Mpc from their centres.}
\label{fig:mass_ratio}
\end{figure}

In this subsection we explain how we compute masses and typical radii of the main large-scale structures to compare their physical properties. When the full three-dimensional velocity field $\boldsymbol{v}(\boldsymbol{r})$ is available, such as from detailed cosmological simulations or, in this case, reconstructed from observational peculiar velocity datasets, it becomes possible to estimate the mass enclosed within a spherical volume of radius $R$ using the principles of linear theory. The process begins by calculating the divergence of the velocity field,  which quantifies how much the velocity field converges or diverges at each point in space. This divergence is then related to the density contrast $\delta (\boldsymbol{r})$ through the linear continuity equation, expressed as 
\begin{equation}
    \delta(\boldsymbol{r}) = - \frac{\boldsymbol{\nabla} \cdot \boldsymbol{v}(\boldsymbol{r})}{f H} ,
\end{equation}
where $f$ is the linear growth rate of cosmic structures and $H$ is the Hubble Constant. By integrating this density contrast over the spherical volume, specifically summing contributions within the sphere, the total enclosed mass can be estimated, such that:
\begin{equation}
    M(<R) = \bar{\rho} \int_{|\boldsymbol{r}| < R} (1 + \delta(\boldsymbol{r})) \, d^3r .\
\end{equation}
As the data are discretized into voxels, this integral is performed numerically by summing over all voxels within the sphere, multiplying the volume of each voxel by the local density contrast. Here, $\bar{\rho}$ represents the mean matter density of the Universe, serving as a baseline for the density fluctuations, with $$\bar{\rho} = \frac{3H_0^2\Omega_m}{8\pi G}\approx 4.63 \times10^{10}~\text{M}_\odot~ \text{Mpc}^{-3}. 
$$

To accurately calculate the mass of these structures, we first need to isolate each individual supercluster region. We use the density grid to identify all local maxima. We impose a minimum spatial separation of 20 \hmpc\ between any two maxima to enforce realistic separations between supercluster domains. Additionally, a minimum threshold of the density contrast between adjacent local maxima is prescribed. These requirements ensure that each maximum represents a genuinely distinct physical region, preventing us from counting multiple peaks that merely trace the same underlying structure. 

Once these maxima have been selected, we use each one as a seed for a flood-fill algorithm of the grid: at each step of this procedure, the regions associated with each seed are expanded outward by one layer of neighbouring grid cells, progressively "filling" the superclusters starting from their central peaks. As the regions grow, we assign and maintain unique group identifiers for each supercluster, so that the emerging structures remain clearly distinguished throughout the process.

\begin{figure*}[h!]  
\centering
\includegraphics[width=\linewidth,angle=-0]{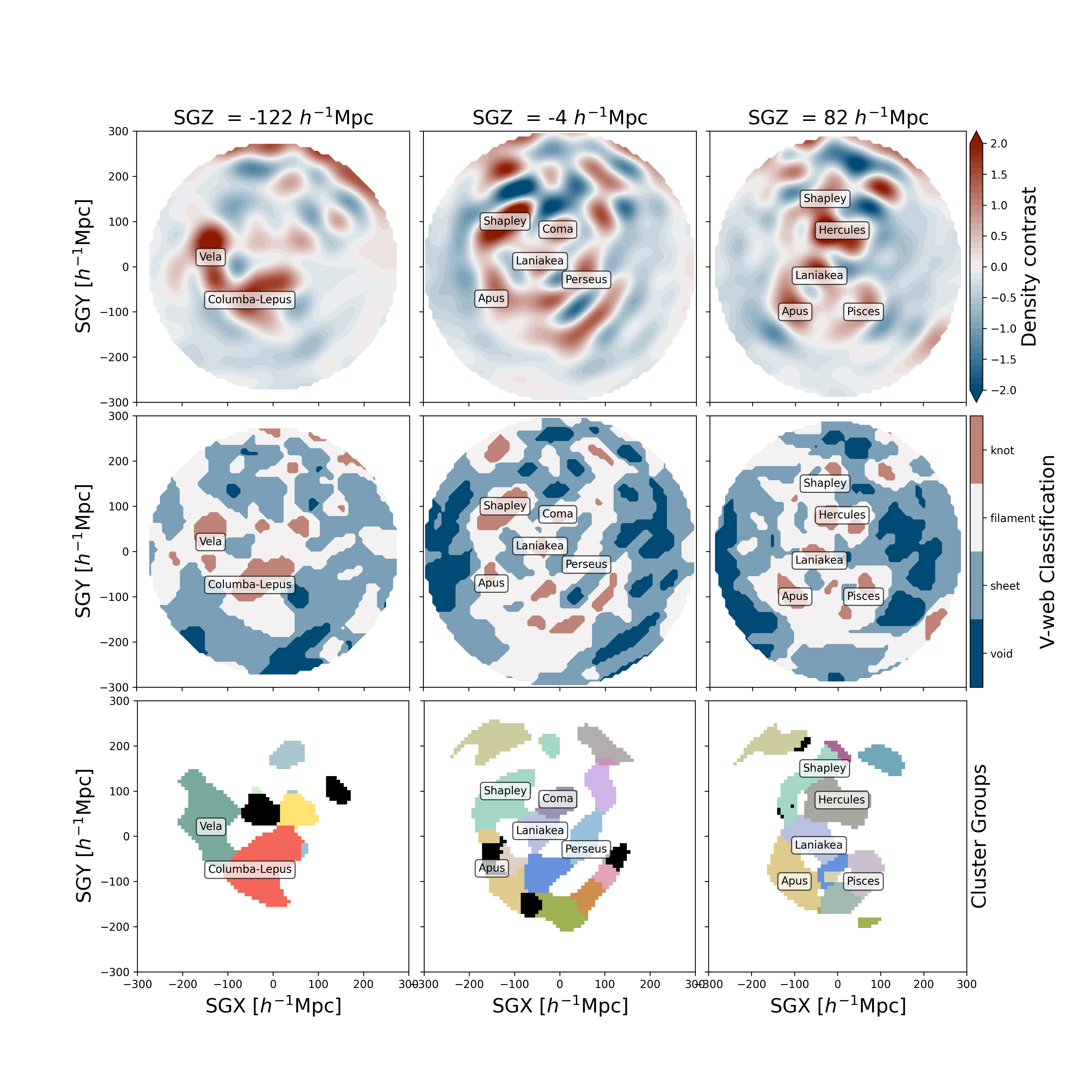}
\caption{Comparison of the density contrast (top row), the V-web classification (middle row), and the final structure groupings (bottom row). The left column shows a slice through  SGZ -122 \hmpc, revealing the prominence of  the Vela and Columba-Lepus superclusters; in the plane centred on SGZ=4 \hmpc\ (middle column) Shapley, Coma, and Apus are most prominent; while the plane centred on SGZ=82 \hmpc\ (right column) features Hercules, Pisces and Laniakea. The black structures in the final row show where clusters were identified but discarded from the final catalogue for not meeting the selection criteria. } 
\label{fig:vweb}
\end{figure*}

Figure \ref{fig:vweb} provides a comparative visualization between different methods of cosmic structure classification. The top panels display the density contrast maps, highlighting regions of varying matter density. The middle panels show the V-web classification, which delineates the cosmic web into voids, sheets, filaments, and knots based on velocity shear. To calculate the V-web, we systematically identify voids and knots by analysing the eigenvalues of the velocity shear tensor across a range of thresholds. We additionally exclude structures that are ill-defined using the full ensemble of the CF4++ZOA HMC velocity reconstructions. For further details, we refer the reader to Section~2 of \cite{HollingerVweb2026}. The bottom panels present the final structure groupings, identifying the prominent superclusters using the method outlined above with the additional constraint that adjoining clusters are presented as a singular structure if connected by $\delta>1.5$. Vela, which forms part of the structures in the left column, consistently exhibits a large density contrast with its double-core morphology clearly evident in the V-web classification. %with a split into two cores clearly evident when classified using the V-web technique.

\begin{table*}[!h]
    \caption{Details of eight Local Universe superclusters, ordered by decreasing central mass. Galactic coordinates and distance, Supergalactic cartesian coordinates, mass, volume, escape velocity and Hubble expansion velocity of eight Local Universe superclusters, ordered by decreasing central mass. The coordinates (SGX,SGY,SGZ) correspond to the highest peak location in the density contrast field and in cosmic V-web for the values in brackets. The masses listed in the table assume $H_0=74.6$, the volume is just the number of voxels $\times (1000/128)^3$. Mass and volume values correspond to the full volume enclosed within the iso-contour of homogeneity ($\delta =0$). The values in brackets correspond to the centre of the supercluster for which the local density contrast values are above 1.5. Vela is a close second in mass and extent to Shapley. It is similar to Hercules in mass but much more extended. None of the superclusters are gravitationally bound at their outermost boundaries where the typical difference between the gravitational escape velocity and the Hubble space expansion velocity is of the order of 2000 \kms. Close to their centres, within the iso-contour of $\delta = 1.5$ that difference is more balanced at $\sim$ 250 \kms.}
    \centering
    \begin{tabular}{lccrrrr}
    \hline
 & $\ell$,$b$, Distance & SGX,SGY,SGZ   & Mass &Volume & Escape vel. & Hubble exp. \\
 & \degr, \degr , $h_{100}^{-1}\text{Mpc}$  & $h_{100}^{-1}$Mpc & $10^{16}$ M$_\odot$& $10^5 h_{100}^{-3}\text{Mpc}^3$ & \kms & \kms\\
 \hline
  Shapley &312 , 48 , 149 & -98,113,4 (-106, 114, 3) & 39.1 (10.1) & 20.2 (3.0) & 5657 (3951) & 7843 (4154) \\ 
   Vela & 266 , 1 , 189 & -121 , -20 , -145 (-122, -11, -161) &  30.6 (7.0) & 16.8 (2.1) & 5160 (3490) & 7376 (3688)\\ 
   Hercules & 27 , 46 , 113 & -27,74,82 (-27, 74, 82) & 17.0 (5.8) & 7.9 (1.7) & 4362 (3291) & 5736 (3437)\\ 
    Columba-Lepus & 250 , -33 , 157 &-51,-74,-129  (-59, -74, -122)  &35.5 (4.4) & 19.6 (1.5) & 5417 (2927) & 7765 (3297) \\ 
    Apus &350 , -35 , 155 & -105 , -98 , 59 (-106, -98, 66) & 31.1 (4.0) & 17.2 (1.3) & 5182 (2858) & 7434 (3143) \\ 
      Pisces  & 98, -44, 133 &74, -98, 51 (74, -89, 59) & 16.7 (2.9) & 8.6 (1.0) & 4262 (2542) & 5900 (2880) \\ 
   Perseus &178 , -39 , 102 & 59,-59,-59 (66, -51, -59) &  18.9 (1.9) & 10.0 (0.7) & 4422 (2184) & 6205 (2557) \\ 
    Laniakea &332 , 6 , 44 & -43,-12,74 (-43, -3, 82)& 15.7 (1.5) & 9.0 (0.5) & 4101 (2052) & 5990 (2286)\\ 
 Coma & 265 , 85 , 82 & -4,82,4  (20, 82, -27) & 6.2 (0.9) & 3.3 (0.3) & 3047 (1731) & 4288 (1928) \\ 
 \hline
    \end{tabular}
    \label{tab:volume-mass}
\end{table*}

Even though superclusters are not spherical, their integrated volumes can be used to compare typical radii. If we were to assume they were spheres, Vela has a typical radius of 74 \hmpc,  Shapley 78 \hmpc\, and Laniakea 60 \hmpc\ (see Table~\ref{tab:volume-mass}). 
An alternative way to define the characteristic radius of a supercluster is to determine the distance from its centre at which the density field reaches the mean cosmic density, i.e. where $\delta = 0$.
Figure~\ref{fig:mass_ratio} shows the mean mass density profiles of nearby superclusters. While the majority of these superclusters attain the average cosmic density at distances of $\sim$70--80 \h Mpc from their centres, they have significantly different density profiles. The Coma cluster, or supercluster, is the smallest among the local superclusters. Both the Shapley and Hercules superclusters display steep density profiles, indicating a rapid increase in density toward their cores. In contrast, the Vela supercluster is significantly more extended, with an estimated radius of about 100 \h Mpc. Vela exhibits a more gradual, shallower decline in density compared to the other large superclusters. This can be interpreted as a different formation history, with internal merging still ongoing. 

As reported in Table~\ref{tab:volume-mass}, Vela encompasses a total mass of $30.6 \times 10^{16}$ M$_\odot$, which is close to the mass of Shapley (78\%) and almost twice the mass of Laniakea. The escape velocity of a sphere of volume $16.8 \times 10^5$(\hmpc)$^3$ and mass $30.6 \times10^{16}$\Msol \ is 3424 \kms. At this radius, the Hubble expansion velocity for Vela is about 7574 \kms assuming $H_{0}$ = 74.6 \kmsmpc. Thus, at the furthest edge of Vela, such an idealized sphere's cosmic expansion dominates over internal gravity (see Appendix \ref{app:A} for the details on computing these velocities in the proper co-moving units). We can also compute these velocities for the central region of Vela within the iso-contour at $\delta=$1.5. In this central region, expansion and gravitation are almost balancing each other. Such an idealized object would be gravitationally bound in this simplified Newtonian picture. Shapley presents a similar case of a large-scale structure that is largely expanding at its furthest iso-contour, while it is almost bound within the central iso-contour $\delta=$1.5, at the current age of the Universe. The inner part of Laniakea and Coma, the smallest local superclusters, are also close to the turnaround radius. 
We additionally tested the method using a full 300 Mpc $h^{-1}$ MDPL2 mock catalogue. For the examined simulated structures, the masses estimated within the defined "supercluster region" -- using the density contrast and the flood-fill algorithm -- agreed with the true halo masses to within approximately 10\%, which we thus take as our error bar for all masses and volumes in Table~\ref{tab:volume-mass}.

We also find that the typical supercluster masses enclosed within the $\delta=0$ iso-contour are nearly an order of magnitude larger than those reported in the review by \cite{Einasto2025} and references therein, but the masses enclosed within the iso-contour  $\delta > 1.5$ are comparable. 
This discrepancy probably arises from methodological differences since most existing supercluster mass estimates rely primarily on luminosity-based methods rather than analyses of the underlying density or velocity fields. The discrepancies could also stem from differences in the definition of the system's spatial extent, and therefore its total mass. For example, Shapley's frequently cited value of  $5 \times10^{16}$ M$_\odot$, attributed to \cite{Proust2006}, appears to be based on an order-of-magnitude estimate. Despite these methodological differences, the two approaches yield results that are broadly consistent and mutually supportive, since we find $10.1 \times10^{16}$ M$_\odot$ for the central region of Shapley.

\subsection{Bulk flow}

\begin{figure}	 
\centering
 \includegraphics[width=\linewidth,angle=-0]{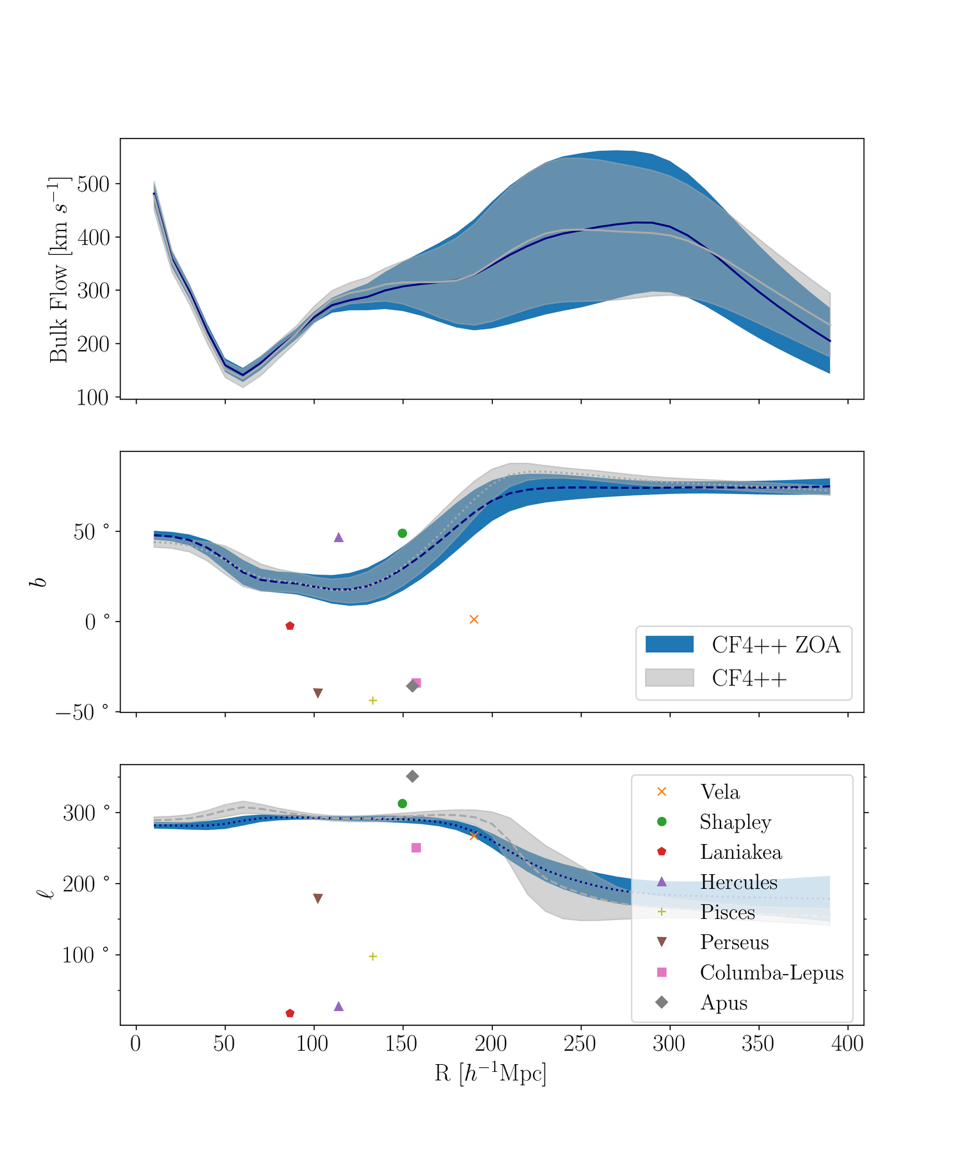}
 \caption{Comparison of the amplitude (top) and Galactic $b$ (middle) and $\ell$ (bottom) directions of the Local Universe bulk flow before (gray) and after adding the ZOA (blue) datasets.  Mean and standard deviation (transparent bands) are calculated using the nearly 10,000 HMC realizations. Galactic positions of the superclusters can be compared to the bulk flow direction.}
 \label{fig:Bulk}
 \end{figure}

Figure~\ref{fig:Bulk} presents a comparison of the amplitude (top panel) and direction (middle and bottom panels) of Local Universe bulk flow before (gray) and after adding the ZOA (blue) datasets. The CF4++ZOA reconstruction of the Local Universe velocity field is better constrained in Galactic longitude ($\ell$): that is to say, the variance is smaller. %The direction of the bulk flow corresponds to the direction of Shapley and Vela at their respective distance,
The direction of the bulk flow aligns with the directions of the Shapley and Vela superclusters at their respective distances, confirming that they are the main actors driving the local gravitational velocity field.

Compared to our previous reconstruction using CF4++ \citep{Courtois2025}, we see no significant changes in the amplitude of the bulk flow (see top panel of Fig.\ref{fig:Bulk}), which is well constrained out to 250 \hmpc\  with an amplitude of $\sim$400 \kms, well above the expected value if the scale of homogeneity was reached at that distance (90 \kms\ in a $\Lambda$CDM prediction).
This measurement is consistent with other measurements of the Local Universe inhomogeneity such as the DESI DR1 conditional density \citep{Sylos2025}.

Similarly, there are minimal changes in the bulk flow direction in terms of Galactic latitude as a function of distance. However, we do see modest shifts in the Galactic longitude direction of the bulk flow at $\sim50$ \h Mpc and $\sim190$ \h Mpc. The bulk flow now remains almost constant at $\ell \approx 280 \degr$ out to 175 \h Mpc.

\section{Conclusions}\label{sec:conclusion}
This article reveals the double-core Vela supercluster, which is almost as massive as Shapley and approximately twice as massive as Laniakea. It is very extended, with a radius of 74 \hmpc\ (assuming a spherical radius for its mass) and suggests a connection to the lower redshift Columba-Lepus supercluster located below the ZOA. All local major superclusters in the Local Universe have expansion velocities larger than the gravitational escape velocity at their outermost boundaries, but they are close to being gravitationally bound within the iso-surface corresponding to a density contrast of $\delta > 1.5$. The updated reconstruction of the velocity and density fields presented here delivers a better characterization of the regions around Norma but results only in marginal changes to the Local Void, the Great Attractor, and Ophiuchus. The addition of these southern hemisphere data does not reduce the bulk flow, which remains at about 400~\kms \ at 250 \hmpc\  distance, which exceeds three times the amplitude that would be predicted for a $\Lambda$CDM universe (90 \kms), indicative of the bulk flow tension as discussed in works such as  \cite{Bouillot2014}.

Having demonstrated that this hybrid redshift \& peculiar-velocity approach can be applied successfully to several thousand galaxies in one of the most challenging regions of the sky, the next step is to extend the methodology to the vastly larger samples that are now becoming available. 

Surveys such as DESI, 4MOST/4HS, and WALLABY will provide hundreds of thousands of redshifts alongside similarly large sets of peculiar-velocity measurements, and fully harnessing these datasets will require reconstruction techniques capable of handling their scale and complexity. The work presented here shows that such an integrated framework is feasible; the task ahead is to generalize and refine it so that it can operate efficiently and reliably on the much richer data streams soon to be delivered by the new generation of cosmological surveys.

\section*{Data Availability}
All the datasets that were used are already published by the observers, apart from the private communication ones that will be published separately by Kraan-Korteweg et al.. The reconstructed density and velocity fields are available as usual:  at the website of H.~M. Courtois \hyperlink{Cosmic-Flows}{https://projets.ip2i.in2p3.fr/cosmicflows/} or upon request if specific help or computational resolution is wanted.

The  simulated data used in this  article are publicly available from the COSMOSIM  database \url{https://www.cosmosim.org/}, with their respective publications cited in section \ref{sec:MDPL2}.

\begin{acknowledgements}

 We thank Dr. Jer\^ome Leca from RSA Cosmos company in France for creating figures ~\ref{fig:superclusters} and ~\ref{fig:superclusters_iso} and their associated animations presented in this article. The authors gratefully acknowledge A. Louw, N. Steyn and S. Kurapati as important contributors to the HI-ZOA datasets.
We express our acknowledgment and respect for the use of the land in South Africa and Australia on which the telescopes that were used are located, and the communities and custodians of these lands; in this context, given the importance of the SARAO Meerkat data at lowest latitude in this endeavour, we introduce a locally inspired affectionate name for the VELA supercluster, "Vela-Banzi", a Xhosa language name meaning "revealing widely".
 HMC acknowledges support from the Institut Universitaire de France. JM acknowledges support from ARC grant CE2100008.
 The MeerKAT telescope is operated by the South African Radio Astronomy Observatory, which is a facility of the NRF, an agency of the Department of Science and Innovation.
 The CosmoSim database used in this paper (MDPL2) is a service by the Leibniz-Institute for Astrophysics Potsdam (AIP). 
 AI-assisted tools were employed to improve some wording and grammar. 
 
\end{acknowledgements}

\bibliographystyle{aa} % style aa.bst
\bibliography{biblio}

\begin{appendix}\label{app:A}

\section{Computing gravitational escape velocity and Hubble expansion in co-moving distances}

This appendix details the calculations employed to estimate the gravitational escape velocity of galaxy superclusters and the Hubble expansion velocity at specified co-moving radii, as summarized in Table 1. These estimates are grounded in the cosmological parameters and voxel-based density fields derived from the CF4++ reconstruction, which encompasses a cubic co-moving volume of $(1000 h^{-1}_{100} \textrm{Mpc})^3$, subdivided into $128^3$ voxels. Consequently, each voxel has a linear size of $\ell = 7.8125\ h^{-1}_{100} \rm{Mpc}  $, corresponding to a volume of $V_{\rm voxel}= \ell^3= 476.8\ h^{-3}_{100} \rm{Mpc}^3$

Adopting a Hubble constant matching that of CF4++, that is $H_0=100h$ \kmsmpc with $h=0.746$, and a matter density parameter $\Omega_m=0.3$, the critical density is calculated as
$$\rho_{\rm crit} = 2.775 \times 10^{11} h^2 M_\odot \mathrm{Mpc}^{-3} \approx 1.54 \times 10^{11}\, M_\odot \mathrm{Mpc}^{-3} \\. $$ The physical size of each voxel is then 10.47 Mpc, with a corresponding volume of $V_{\rm voxel}= 1148.6 \rm{Mpc}^3$.
The mean matter density follows as $\bar{\rho} = \Omega_m \rho_{\rm crit}$, and the mass within a voxel with over-density $\delta_i$ is 
$$ M_i = \bar{\rho}(1+\delta_i) V_{\rm voxel} \approx 1.77 \times 10^{14} \Omega_m (1+\delta_i)\, M_\odot \\. $$

As an example, consider the Shapley supercluster, which spans a volume of  $V = 2.02 \times 10^{6} \,h^{-3}{\rm Mpc}^3$ and has a mass of $3.91 \times 10^{17} \, M_\odot$. Assuming spherical symmetry, its radius can be estimated as $R = (3V/4\pi)^{1/3} \approx 105$ Mpc. 
The gravitational escape velocity at this radius is given by $$ v_{\rm esc} = \sqrt{2GM/R} \approx 5657\ {\rm km\ s}^{-1}\\, $$ where $G = 4.302 \times 10^{-9} {\rm Mpc\ } ({\rm km\ s}^{-1})^2\ M_\odot^{-1}$. The Hubble expansion velocity at this radius is $$v_H = H_0 R \approx 7843\ {\rm km\ s}^{-1} \\. $$
This comparison illustrates the interplay between the gravitational binding of the supercluster and the cosmic expansion at this scale.

\section{Tests}\label{app:tests}

\subsection{MDPL2 mocks}\label{sec:MDPL2}
We used the MultiDark Planck 2 (MDPL2) simulation, which is part of the MultiDark project, a suite of cosmological N-body simulations \citep{klypin_multidark_2016}. All simulations in this suite adopt a flat $\Lambda$CDM cosmology with the following parameters: $\Omega_\Lambda=0.692885$, $\Omega_M=0.307115$, $h=0.6777$, $\sigma\sbr{8,linear}=0.8228$, and $n_s=0.96$, in agreement with the Planck measurements \citep{planck_2020-values}. MDPL2 contains 3840$^3$ dark matter particles, each with mass $1.51\times 10^9$ \h \Msol. Haloes were identified using Rockstar \citep{behroozi_rockstar_2012}, which led to over 10$^8$ haloes. Halo merger trees were subsequently constructed with ConsistentTrees \citep{behroozi_gravitationally_2013}. The simulation volume is a periodic cube of side length 1000 \hmpc. Our analysis is restricted to the $z=0$ snapshot.
The halo catalogue was retrieved from the COSMOSIM database\footnote{\url{www.cosmosim.org}}.

We select a random region from the MDPL2 simulation, ensuring a volume with a radius of 300 \hmpc\ to match the survey depth of CF4++. Within this region, we calculate the radial velocity of halos in the CMB frame with respect to the origin, using the Cartesian peculiar velocities provided. To reflect the limitations inherent to observational surveys, we apply increasingly strict cuts in halo mass as a function of distance from the origin, starting from approximately $\sim 5\times 10^{9}$ \Msol\ and increasing logarithmically up to $10^{12}$ \Msol\ at 300 \hmpc. Additionally, we convert the positions of the halos to Galactic coordinates, and exclude all objects with $|b|> 15^\circ$  to isolate the Galactic plane. The final selection of halos is determined following the criteria outlined in the previous section. 

\subsection{Description of Tests}
To validate our methodology and ensure that we are not over-analysing the dataset, we conduct three types of  tests. 
\begin{itemize}
    \item The first set of tests (\textit{ZOA}) uses the zone of avoidance  C1-C4 samples and CF4++ datasets directly. We use both the catalogues and previous reconstructions to assign peculiar velocities and distance moduli within our HMC reconstruction scheme. By design we expect these reconstructions to reproduce the original CF4++ density and velocity fields.
    \item The second set  (\textit{MDPL2}) involves introducing simulated data into the ZOA. We use the MDPL2 halos directly for these tests as the actual luminosity of these objects is not important for our purposes. To maintain consistency, we also cut the number of simulated data down to include at most an equivalent number of galaxies as in the original sample, allowing us to assess the robustness of our analysis procedures under controlled conditions. As expected, the recovered structure from the HMC reconstruction is coherent and matches the true density field that one would recover in the ZOA region. 
    \item The third set of tests (\textit{random}) produces data that is largely incoherent, with the only consistent aspect being the number of objects in the ZOA sample. In this test, the mock galaxies are randomly distributed within the ZOA, with their positions assigned arbitrarily. Their peculiar velocities are also assigned randomly, aligning with the distribution of the CF4++ sample. This approach results in a dataset that does not preserve the underlying spatial or kinematic correlations observed in the original sample. 
\end{itemize}
In all cases, the uncertainties associated with the generated samples are computed following the methodology outlined in section \ref{sec:test-datasets}.

\begin{figure*}[hbt!]

\begin{subfigure}{.475\linewidth}
  \includegraphics[width=\linewidth]{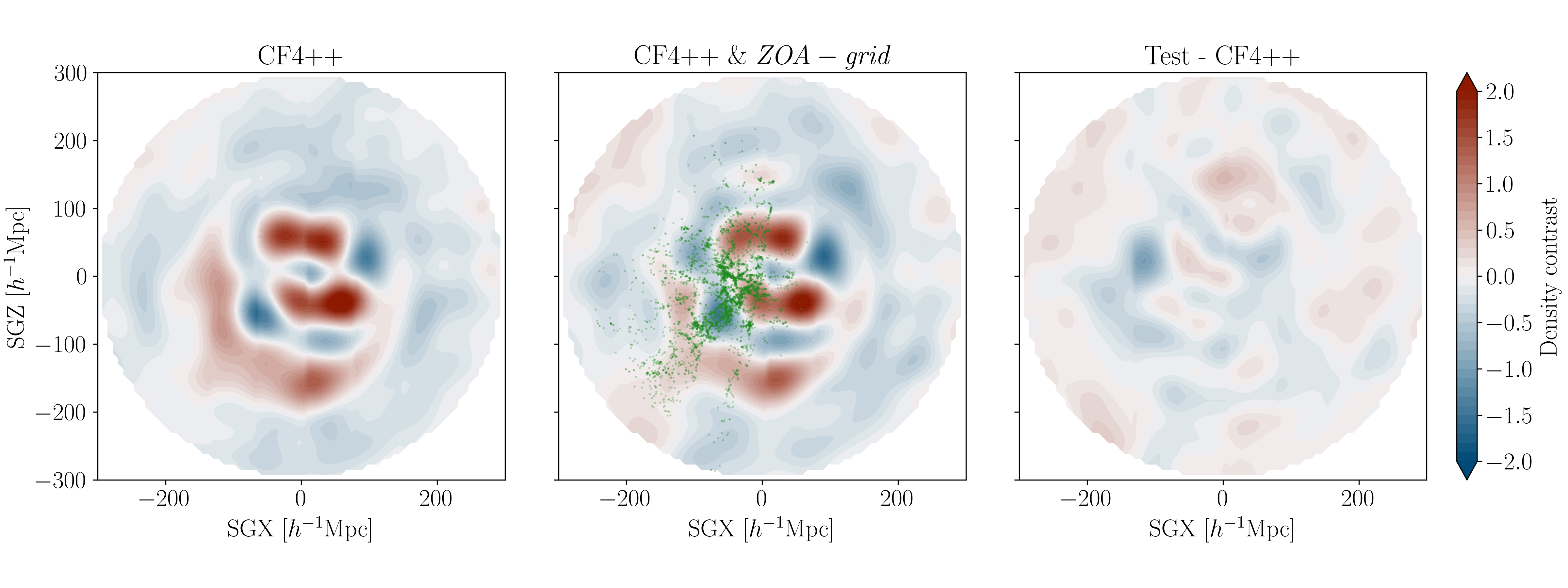} 
\end{subfigure}\hfill 
\begin{subfigure}{.475\linewidth}
  \includegraphics[width=\linewidth]{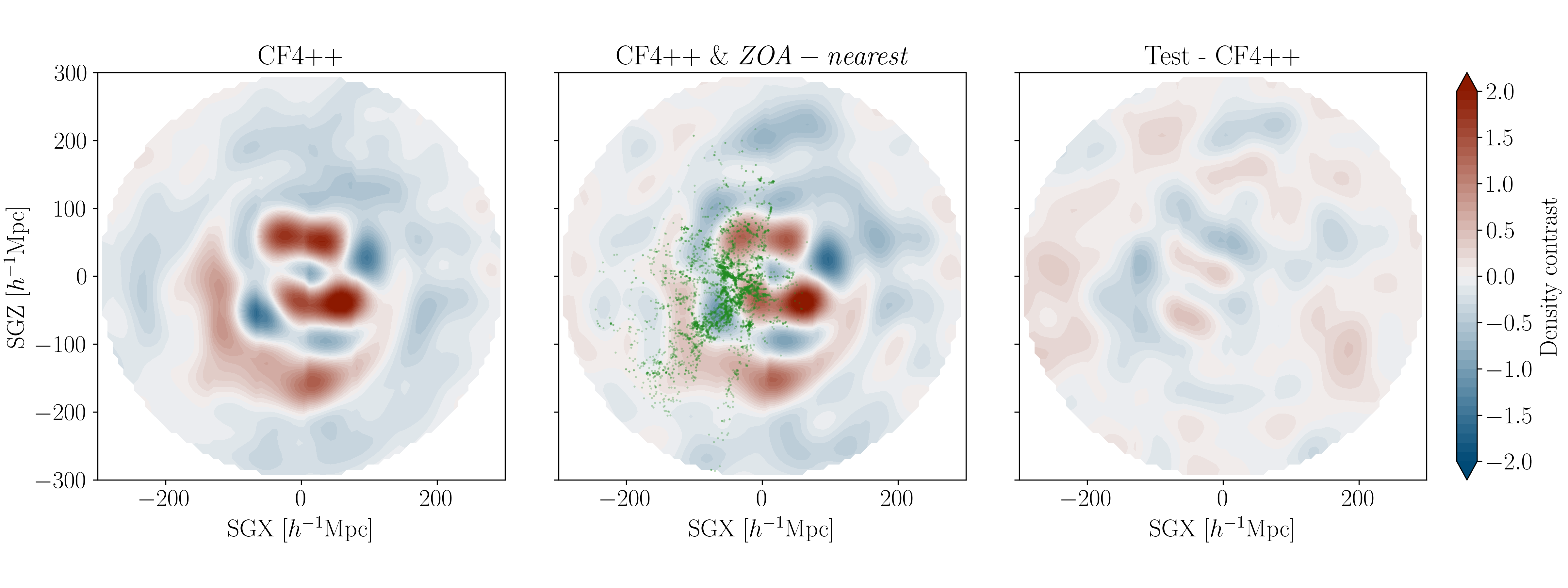} 
\end{subfigure}
\begin{subfigure}{.475\linewidth}
  \includegraphics[width=\linewidth]{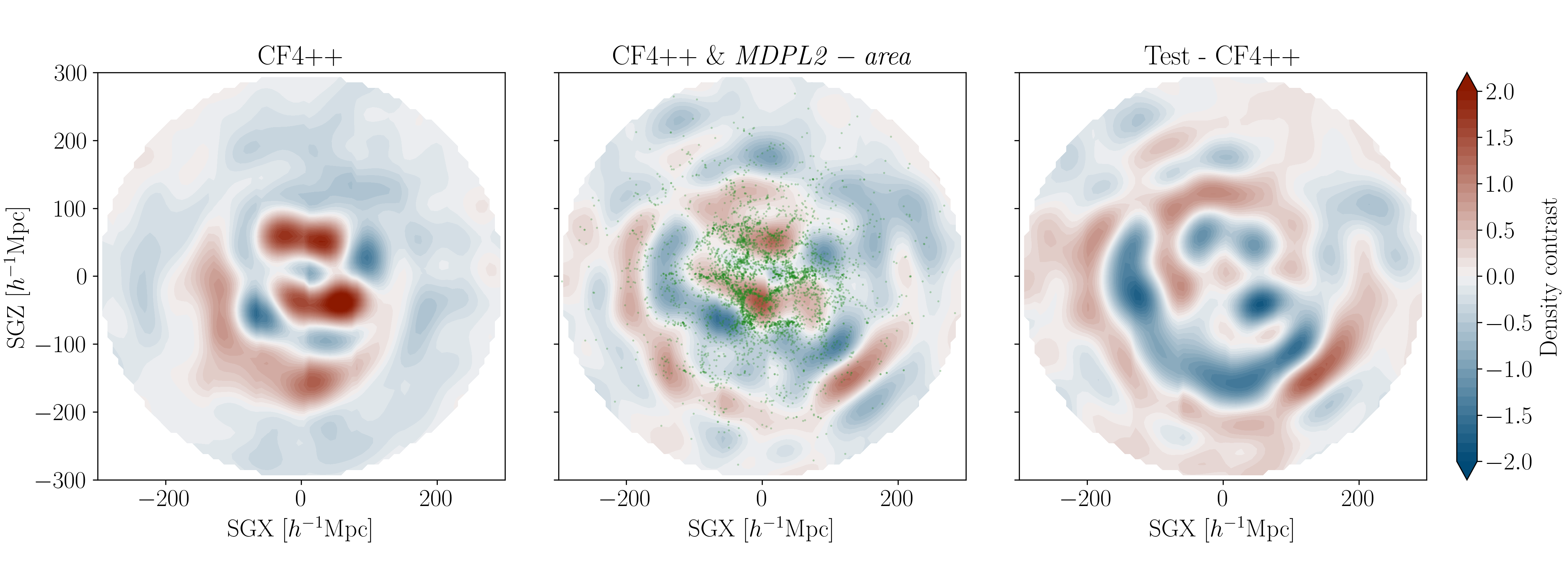} 
\end{subfigure}\hfill  % <-- "\hfill"
\begin{subfigure}{.475\linewidth}
  \includegraphics[width=\linewidth]{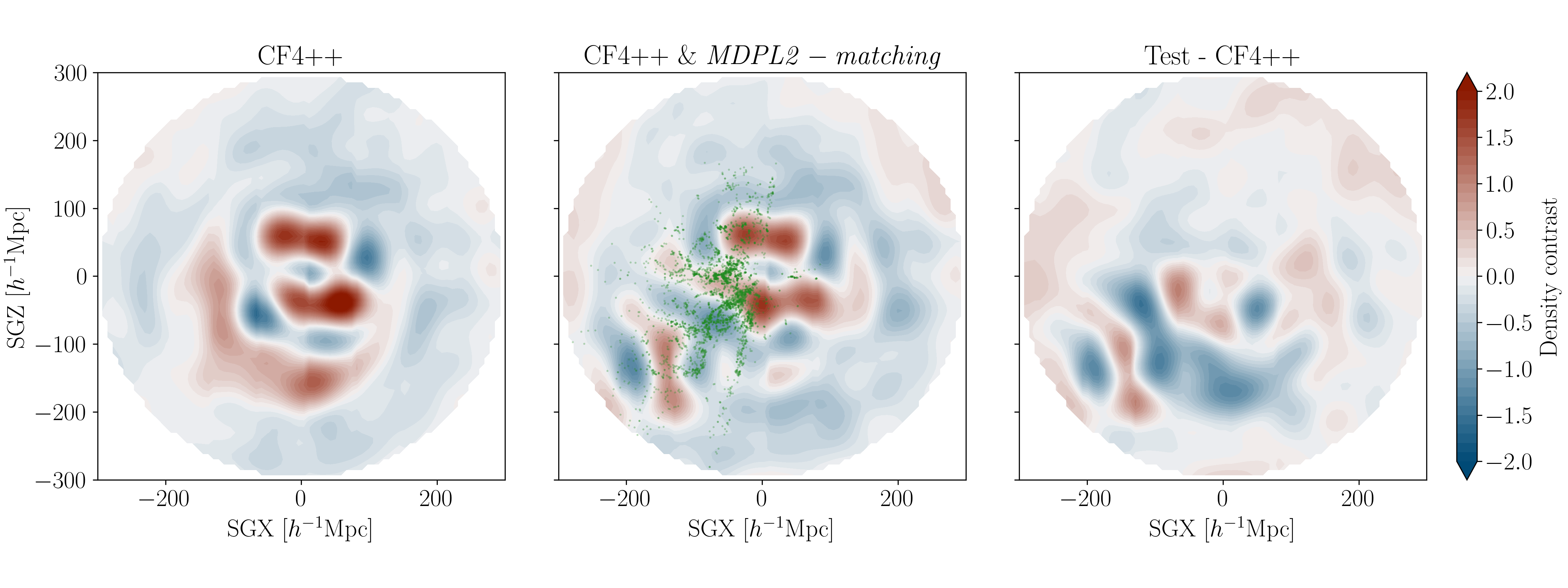} 
\end{subfigure}
\begin{subfigure}{.475\linewidth}
  \includegraphics[width=\linewidth]{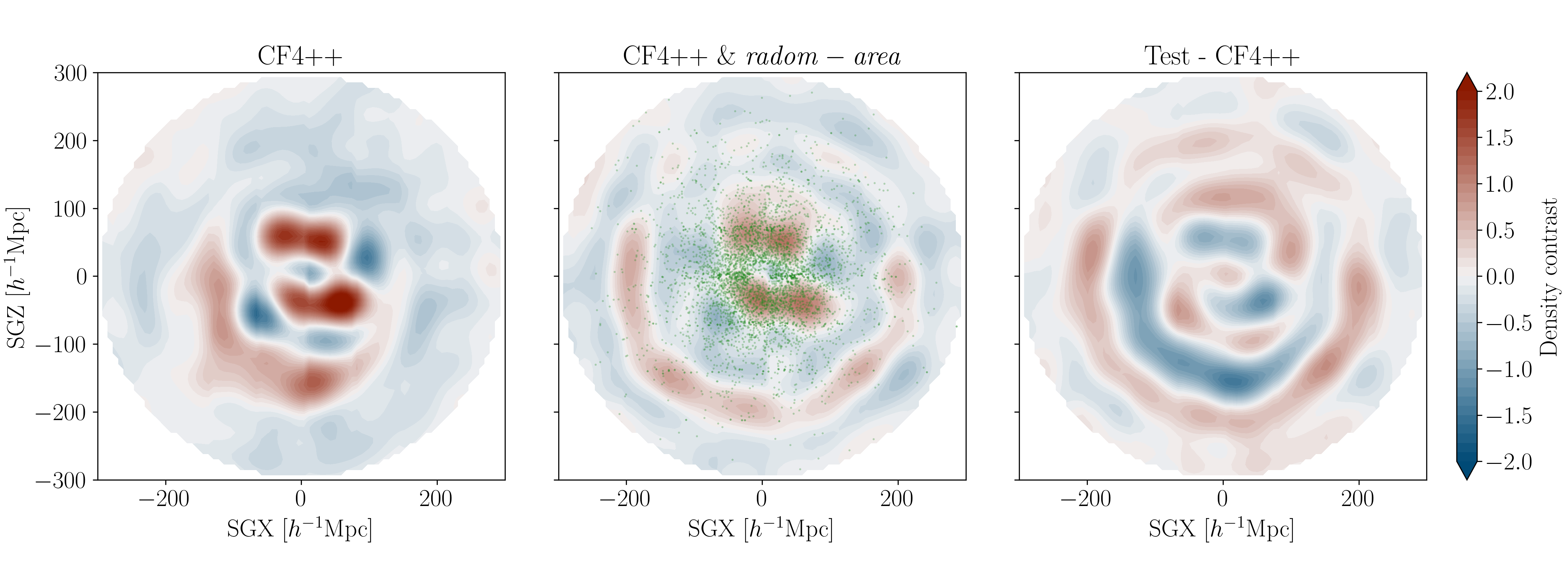} 
\end{subfigure}\hfill % <-- "\hfill"
\begin{subfigure}{.475\linewidth}
  \includegraphics[width=\linewidth]{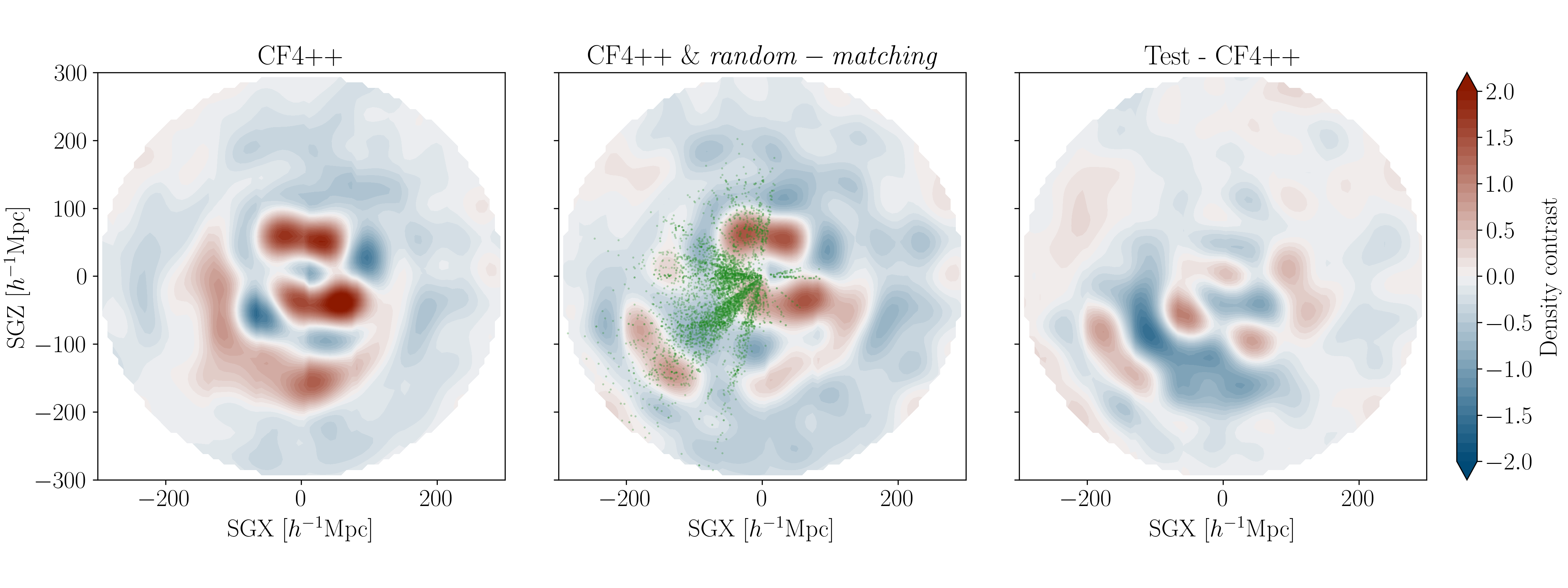} 
\end{subfigure}
\hfill
\begin{subfigure}{0.475\linewidth}\hspace{.5\textwidth}
    \includegraphics[width=\textwidth]{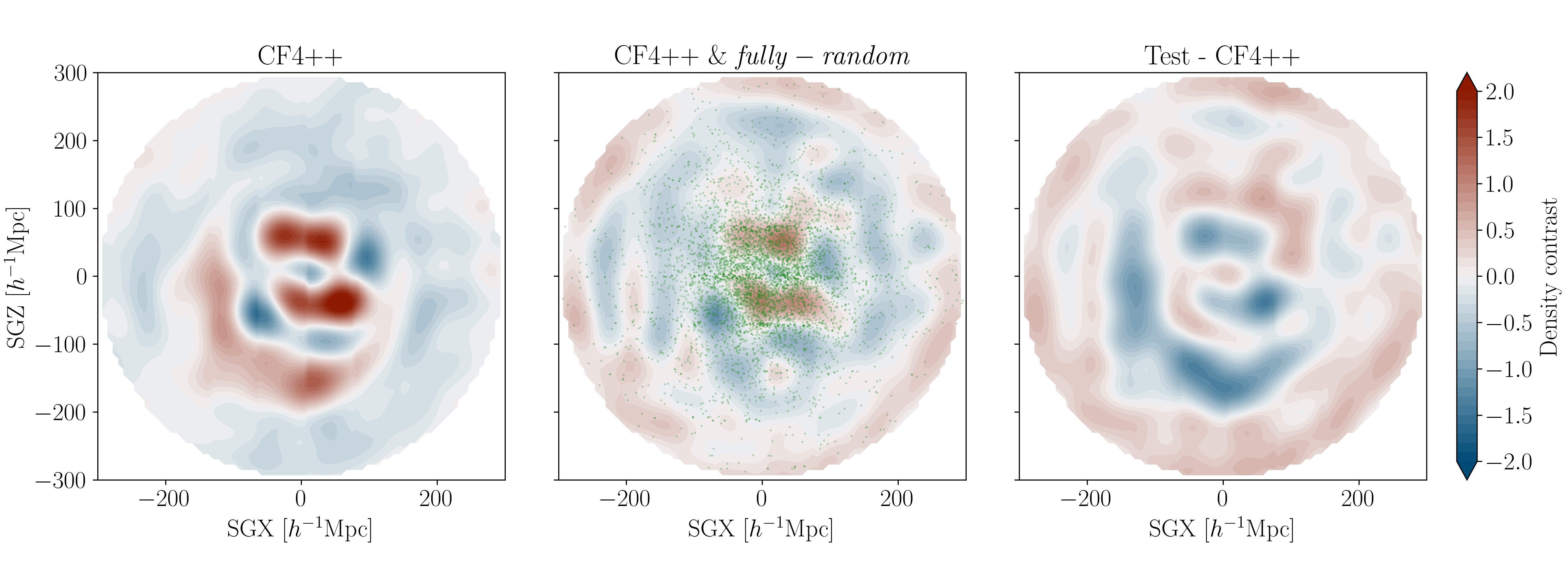}
\end{subfigure}
\caption{The local cosmography sliced through the Galactic plane ($b \approx 0 \degr$), the left and middle columns show the mean of the HMC realizations of the density contrast ($\delta$) for CF4++ and our test cases, respectively. The right column shows the difference between the two. The green points shown in the middle panel demonstrate the positions of the artificial galaxies added in each test case. }
\label{fig:samples}
\end{figure*}

\begin{figure}[h!]
 \centering
\includegraphics[width=\linewidth,angle=-0]{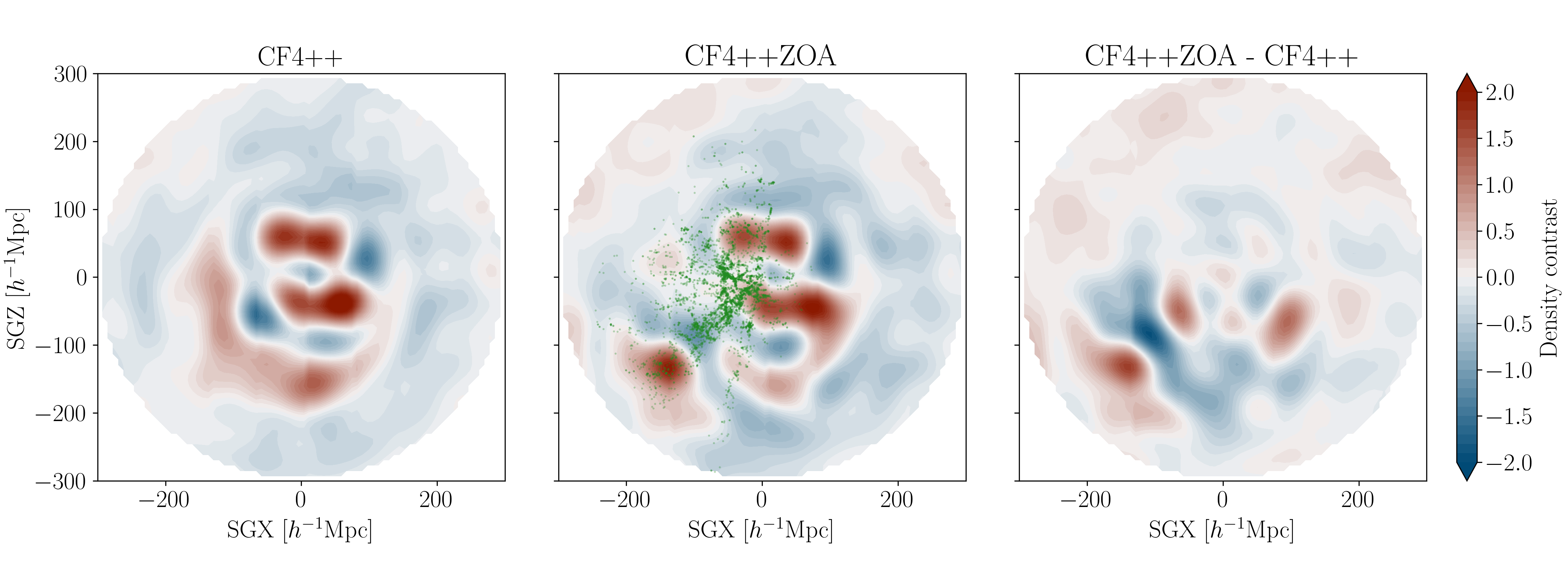} \\
  \caption{Similar to  Fig. \ref{fig:samples} showing the CF4++ZOA reconstruction presented in this work}
\label{fig:ZOAplane}
\end{figure}

Figure~\ref{fig:ZOAplane} shows the changes to the density field through the Galactic plane ($b \approx 0 \degr$), for the CF4++ (left), test cases (middle) and difference between the two (right). As expected, the \textit{ZOA} configurations -- our first tests displayed in the top row -- exhibit the smallest modifications in the reconstructed density field. This behaviour is expected, as  we explicitly constrain the peculiar velocities of the added galaxies to be consistent with our earlier measurements.  

In contrast, in the \textit{fully--random} scenario (fourth row), the density field becomes systematically less extreme across all regions: previously under-dense regions become more populated, while over-dense regions are smoothed out and become less pronounced. As expected, this trend is similarly observed in the \textit{random-area} realizations, and to a lesser extent in the \textit{random-matching} (third row). 

Unsurprisingly, the \textit{MDPL2-area} scenario (second row, left panel) generates the largest changes to the $b\approx0$ plane. While the peculiar velocities assigned to the random case effectively wipe out existing structure. The MDPL2 mock local universe has its own structures that do not directly correspond to our observed Local Universe. This introduces an additional mismatch when used to fill in missing data. The \textit{MDPL2-matching} case also shows this general tendency  (second row, right panel), but to a lesser extent. At the same time, it produces the largest overall discrepancies in the density field, reflecting the fact that the MDPL2 mock catalogues are not an exact representation of the actual Local Universe and thus introduce additional mismatch when used to fill in the missing data.

In all cases, none of the test scenarios exhibit density peaks in the Vela region that are as prominent or extensive as those observed in the CF4++ and CF4++ZOA case, as illustrated in Fig.~\ref{fig:ZOAplane}.

\section{Reconstructions of other structures}\label{app:C}

We analysed 31 well-known large-scale structures in the Local Universe: Antlia, Great Attractor, Lepus, Pisces, Aquarius, Great-Arch, Local Void, Sculptor, Arch-knot,  Shapley, Bootes, Great-Wall, Norma, southern-Wall, Centaurus, Hercules, Ophiuchus, Ursa-Major, Columba,  Pavo-Indus, Vela, Coma, Horologium-Reticulum, Perseus-Pisces, Virgo, Corona Borealis, Hydra,  Phoenix, Fornax, Leo, Pisces-Cetus. The majority of them show minimal to no changes in both their density and velocity fields. 

\noindent{\bf Norma, Great Attractor, Local Void, and Ophiuchus: } 

\begin{figure}[h!] 
 \centering
 \includegraphics[width=\linewidth,angle=-0]{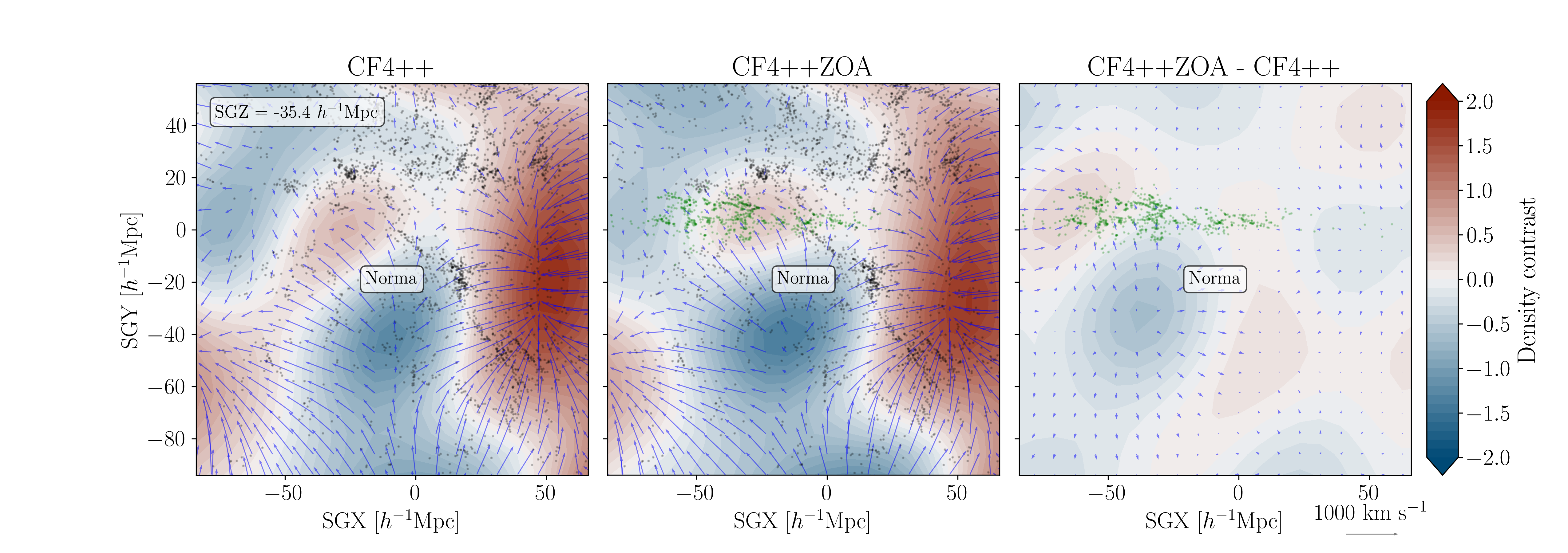} \\
\includegraphics[width=\linewidth,angle=-0]{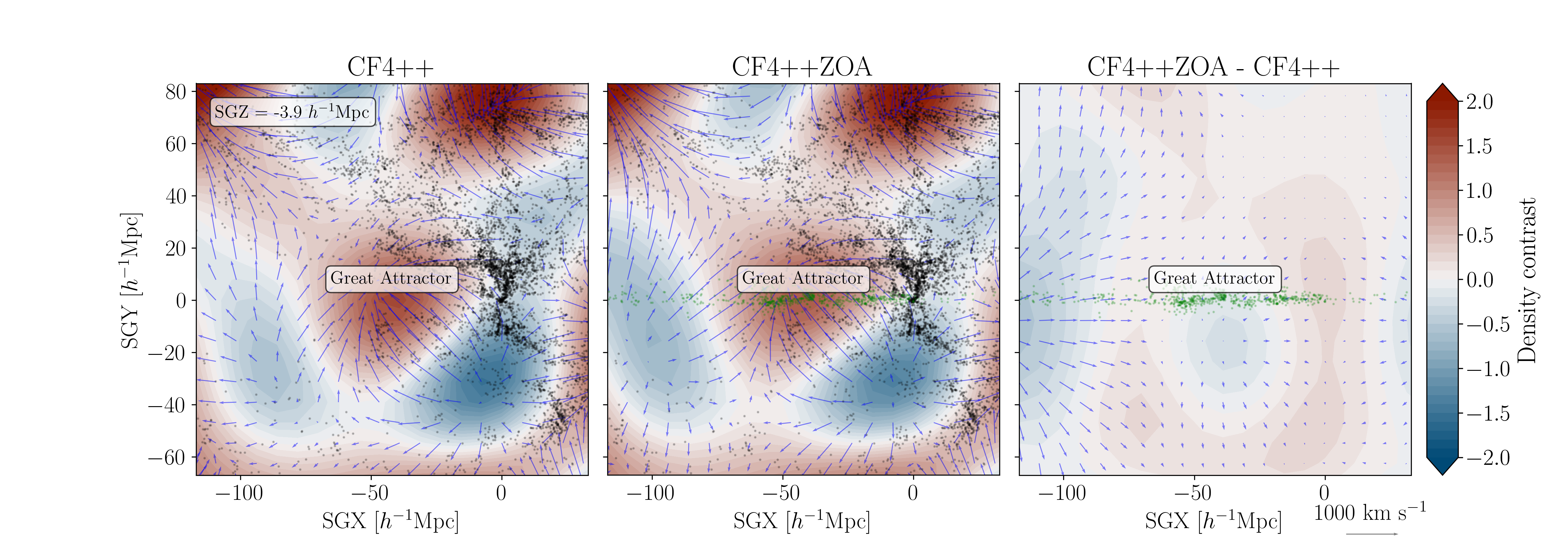} \\
\includegraphics[width=\linewidth,angle=-0]{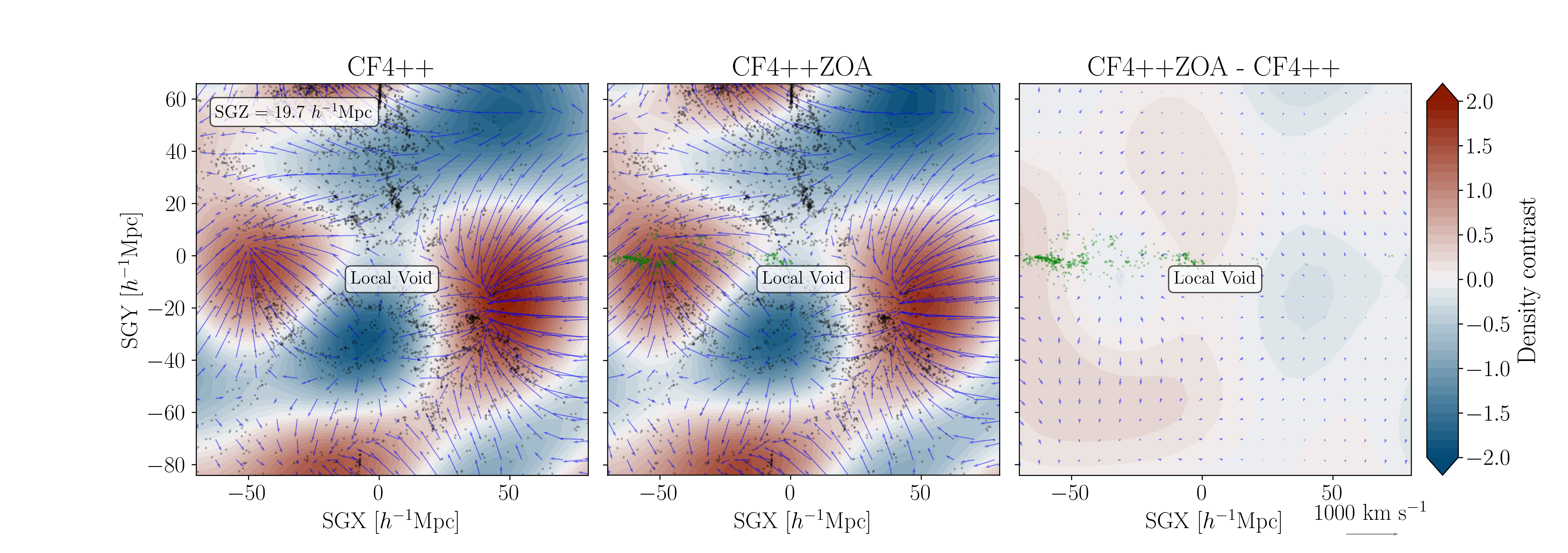} \\
\includegraphics[width=\linewidth,angle=-0]{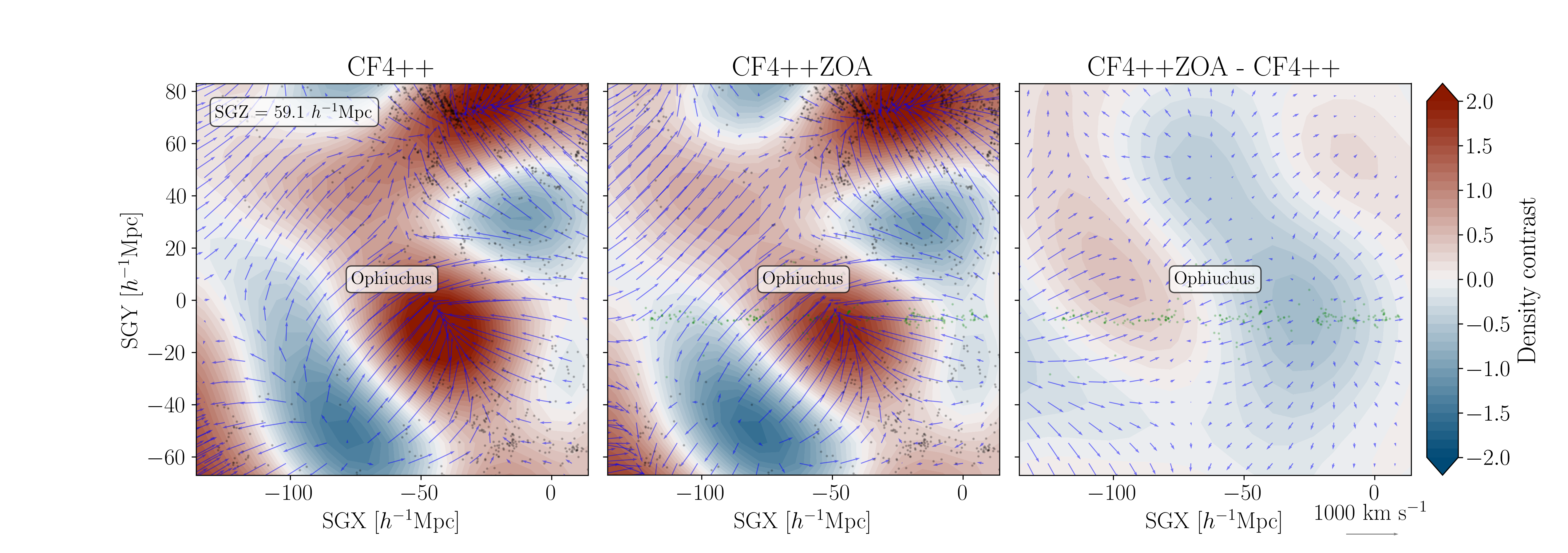}\\
  \caption{Similar to Fig. 3, the reconstructed matter density field, averaged over a total width of 23.4 \h Mpc in the SGZ plane. The left panel displays results derived from the CF4++ dataset, while the middle panel uses the CF4++ZOA dataset, and the right panel shows the difference between the two reconstructions. Overlaid is the mean velocity field. Black dots represent CF4++ galaxies, and the green dots correspond to newly added ZOA data points located within the averaged region. The plots show four structures located near the Galactic plane that interestingly demonstrate no significant changes to their local cosmography despite the addition of the ZOA datasets.  From top to bottom, these are Norma, Great Attractor, the Local Void and Ophiuchus.}
 \label{fig:nochange}
\end{figure}

Figure~\ref{fig:nochange} shows four local large-scale structures that were most likely to be affected by the inclusion of ZOA data due to their proximity to the Galactic plane:  Norma, the Great Attractor, the Local Void and Ophiuchus. Interestingly, no dramatic change is noticeable in their local cosmography, apart from Norma, whose shape is better defined with the inclusion of the ZOA datasets (see top row in Figure C1). It is separated from neighbouring structures, exhibits a round morphology, and is well embedded within the zone of avoidance. Norma appears to lie along the wall of a cosmic void located just below it at (-12;-50) \hmpc\ in SGX-SGY. Although the lack of data immediately beyond the Norma cluster has already been noted%While a lack of data just behind the Norma cluster was already notable 
in \citep{Woudt2004, Woudt2008}, the actual extent --and roundness -- of the void could not be defined with the previously existing data. 

%fig11

\end{appendix}

\end{document}